\documentclass[aps,prd,reprint,superscriptaddress,nofootinbib]{revtex4-2}
\pdfoutput=1

\usepackage[utf8]{inputenc}

\usepackage{amsmath}
\usepackage[pdftex]{graphicx}

\usepackage{amssymb} 

\usepackage{subcaption}

\usepackage{xspace}
\newcommand*{\ie}{i.e., }
\newcommand*{\eg}{e.g., }
\newcommand*{\fig}{Fig.\@\xspace}
\newcommand*{\figs}{Figs.\@\xspace}
\newcommand*{\Eq}{Eq.\@\xspace}
\newcommand*{\Eqs}{Eqs.\@\xspace}
\usepackage{braket}

\begin{document}

\title{Black Hole Metamorphosis   
\\ 
and Stabilization by Memory Burden}

\author{Gia Dvali}
\email[]{georgi.dvali@physik.uni-muenchen.de}
\affiliation{Arnold Sommerfeld Center, Ludwig-Maximilians-Universit\"at, \mbox{Theresienstraße 37, 80333 M\"unchen, Germany}}
\affiliation{Max-Planck-Institut für Physik, F\"ohringer Ring 6, 80805 M\"unchen, Germany}

\author{Lukas Eisemann}
\email[]{lukas.eisemann@physik.uni-muenchen.de}
\affiliation{Arnold Sommerfeld Center, Ludwig-Maximilians-Universit\"at, \mbox{Theresienstraße 37, 80333 M\"unchen, Germany}}
\affiliation{Max-Planck-Institut für Physik, F\"ohringer Ring 6, 80805 M\"unchen, Germany}

\author{Marco Michel}
\email[]{marco.michel@physik.uni-muenchen.de}
\affiliation{Arnold Sommerfeld Center, Ludwig-Maximilians-Universit\"at, \mbox{Theresienstraße 37, 80333 M\"unchen, Germany}}
\affiliation{Max-Planck-Institut für Physik, F\"ohringer Ring 6, 80805 M\"unchen, Germany}

\author{Sebastian Zell}
\email[]{sebastian.zell@epfl.ch}
\affiliation{Institute of Physics, Laboratory for Particle Physics and Cosmology, \'Ecole Polytechnique F\'ed\'erale de Lausanne, CH-1015 Lausanne, Switzerland}
\affiliation{Arnold Sommerfeld Center, Ludwig-Maximilians-Universit\"at, \mbox{Theresienstraße 37, 80333 M\"unchen, Germany}}
\affiliation{Max-Planck-Institut für Physik, F\"ohringer Ring 6, 80805 M\"unchen, Germany}

\date{\today}

\begin{abstract}
	Systems of enhanced memory capacity are subjected to a universal effect of {\it memory burden}, which suppresses their decay. In this paper, we study a prototype model to show that memory burden can be overcome by rewriting stored quantum information from one set of degrees of freedom to another one. However, due to a suppressed rate of rewriting, the evolution becomes extremely slow compared to the initial stage.
	Applied to black holes, this predicts a metamorphosis, including a drastic deviation from Hawking evaporation, at the latest after losing half of 
the mass. This raises a 
		tantalizing question about the fate of a black hole.
		As two likely options, it can  either become extremely long lived 
	or decay via a new classical instability into gravitational lumps. 
	The first option would
	open up a new window for small primordial black holes as viable dark matter candidates.
\end{abstract}

\maketitle

\tableofcontents

\section{Introduction}

\subsection{Big Picture}
This paper is about understanding a very general
phenomenon \cite{1810.02336, 1812.08749}, called {\it memory burden,} exhibited by systems that achieve a  {\it high capacity of memory storage} and its potential applications 
to black holes. The essence of the story is that 
a high load of quantum information carried by such a system   
stabilizes it. This means that, in order to decay, the system must off-load 
the memory pattern from one set of modes into another one. Our studies 
show that this becomes harder and harder with larger size. As a result,  
the quantum information stored in the memory back-reacts and stabilizes the system at the latest after 
its half-decay. The universality of the phenomenon 
suggests its natural application to black holes. 

The present paper is a part of a general program, initiated some time 
ago, which consists of two main directions. 
One is the development of a microscopic theory describing a black hole 
as a bound state of soft gravitons, a so-called quantum $N$-portrait
\cite{1112.3359}.  The occupation number $N_c$  of gravitons is 
critical in the sense that it is the inverse of their (dimensionless) gravitational interaction.
The softness of gravitons here refers to wavelengths 
comparable to the gravitational radius of a black hole.  This criticality
has been identified \cite{1207.4059} as the key reason for understanding the maximal information storage capacity of a black hole
quantified by its Bekenstein-Hawking entropy. 
Due to extremely weak interactions among the constituent gravitons, this framework allows to perform
computations within the validity domain of effective field theory
exploiting the power of $1/N_c$-expansions.  

The second direction  \cite{1207.4059, 1507.02948, 1507.08952, 1601.01329, 1711.09079,1801.03918, 1712.02233, 1805.10292, 1810.02336, 1812.08749, 1906.03530, 1907.07332, 2003.05546}
is to instead use the {\it enhanced capacity of memory storage} as a guiding principle. 
That is, we study generic systems which possess states with a high capacity of memory storage and try to
identify universal phenomena that 
take place  near such states.  The idea then is to come back and apply this 
knowledge to black holes and look for analogous phenomena there. 
The advantage of this approach is that a unique knowledge about 
black holes can be gained by studying systems that are much easier solvable both analytically and numerically. 
The present paper is about the detailed study of one such universal 
phenomenon identified in \cite{1810.02336, 1812.08749}, namely the above mentioned memory burden.  

Before continuing we wish to make a clarifying remark, in order not 
to confuse the reader with our terminology.  We shall often use the term 
{\it enhanced capacity of the memory storage} as opposed 
to, for example, {\it maximal entropy}.
This is because the  former term covers a wider class of systems:
A state can have a sharply 
enhanced memory storage capacity even if the corresponding microstate entropy is not necessarily maximal.  The above studies show that systems that possess such states still exhibit some black hole like properties.  Of course, the converse is in general true:  A state of maximal microstate entropy does possess a maximal memory storage capacity.  In particular, all such states must share the memory burden effect.

\subsection{Main Finding} 
 Let us start with setting the framework.
 Physical systems are characterized by a set of degrees 
 of freedom (modes) and interactions between them.  It is convenient 
 and customary to 
 describe the degrees of freedom as quantum oscillators. 
 The basic quantum states of the system then can be 
 labeled by a sequence of their occupation numbers
 $\ket{n_1,\ldots ,n_K}$.   
 Such a sequence stores quantum information which we can refer 
 to as the {\it memory pattern}.  The efficiency of memory storage is then 
 measured by the number of patterns that can be stored within a certain 
 microscopic energy gap \cite{1801.03918, 1805.10292}.  When this number is high, we shall say that 
 the system has an enhanced capacity of memory storage. 
 The above notion is closely related 
 to microstate entropy but is much more general. If the states describing 
different patterns
 share the same macroscopic characteristics (e.g., the total mass 
 or angular momentum), the microstate entropy can be defined in the usual  way  $S = \ln (n_{\rm st})$, where $n_{\rm st}$ is the number of 
 distinct basic microstates $\ket{n_1,\ldots,n_K}$.

 Naturally, we are interested in systems that dynamically attain
a high capacity of memory storage. This can be achieved if the system reaches a critical state in which a large number of gapless modes emerge. Then information can be encoded in the occupation numbers of the gapless 
modes without energy cost.
 This generic mechanism has already been investigated in a series of papers \cite{1207.4059, 1507.02948, 1711.09079,1801.03918,1712.02233,1805.10292, 1810.02336, 1812.08749}.  Originally, it was introduced 
 for understanding the origin of Bekenstein-Hawking  entropy 
 in a microscopic theory of  black hole's  quantum $N$-portrait
 \cite{1112.3359}. However, it was soon realized in the above 
 papers that this mechanism is universally 
 operative in systems with high capacity of memory storage. 
 Interestingly, it has been repeatedly observed that the information storage
  in such systems exhibits some black hole-like properties.
 
   This universality suggests that by understanding general phenomena and 
applying this understanding to black holes, we can gain new  knowledge that until now has 
 been completely blurred by technical difficulties in quantum gravity  
 computations. Such {\it terra incognita} is the black hole evolution beyond its half evaporation. The reason is that the standard semi-classical computations are unable to account for  quantum back reaction. Therefore, 
they are no longer applicable once back reaction becomes important, \ie the latest by half evaporation.
 Instead, for resolving such questions a microscopic theory is needed
 such as quantum $N$-portrait \cite{1112.3359}.   
 
As mentioned above,  in this microscopic theory the black hole is 
 described as a saturated critical state of soft gravitons of wavelengths 
 given by the gravitational radius of a black hole, $r_g$.  
Since the occupation number is critical, \ie equal to the inverse of their gravitational coupling,
 the kinetic energy of individual gravitons just saturates the
 collective attraction from the rest. As a result, the gravitons form a long-lived bound-state, a black hole.   However, the bound-state 
 is not eternal.  Instead,  due to their quantum re-scattering, 
 the soft bound-state slowly loses its constituents and depletes. 
 On average, it emits a quantum of wavelength $r_g$ per 
 time $r_g$. At the same time, the emissions of quanta that are either much harder or softer 
 are suppressed. In total, this process reproduces Hawking's  evaporation 
up to $1/N_c$ corrections. 
 It is these $1/N_c$-corrections that are responsible for new effects 
 that were invisible in the standard semi-classical treatment. Note that 
 the latter corresponds to the $N_c = \infty$ limit of the microscopic theory.  
 
 The computations performed in the microscopic theory 
  \cite{Dvali:2013eja,Dvali:2012rt,Dvali:2012wq,Dvali:2013vxa,Dvali:2017eba} unambiguously indicate that the classical description breaks down after the black hole has lost on the order of half of its mass, which corresponds to on the order of $N_c/2$ emissions. At this point, the back reaction 
 (i.e., $1/N_c$-effects integrated over time) become
 so important that the true quantum evolution completely 
departs from the naive semi-classical one. 
In this light, the two immediate tasks are: 1) Better quantify the quantum back reaction effects that lead to this breakdown; 
and 2) Predict 
what happens beyond this point.
 
In order to address these questions, 
  we shall try to understand very general 
 aspects of time-evolution of systems of enhanced memory capacity. 
 We shall use the simplest possible prototype model
 with this property. 
  Then we try to 
 extend the obtained knowledge to black holes and  cosmology and speculate 
 about the possible consequences.
The above strategy is the continuation of the one adopted in the previous 
papers   \cite{1207.4059, 1507.02948,1507.08952, 1601.01329, 1711.09079,1801.03918, 1712.02233, 1805.10292, 1810.02336, 1812.08749, 1906.03530, 1907.07332, 2003.05546}. 
The persistent pattern emerging from this work 
 is that systems of enhanced capacity of memory storage exhibit  
 striking similarities with certain black hole 
properties.  For example, they share a slow initial decay via the emission 
of the soft quanta without releasing the stored information for a very long time. In short, it appears to be a promising 
 strategy to try to make progress in understanding 
black holes by abstracting from the geometry and instead viewing their information storage capacity as the key characteristic.

In order to avoid any misunderstanding, we wish to clearly separate 
  solid results from speculations. 
  In the present paper,  we shall focus on a very precise source of quantum back reaction. Following  \cite{1810.02336, 1812.08749}, we shall refer to it as the phenomenon of {\it memory burden}.  The essence of 
  it, as described above, is that the high load of quantum information 
  stored in a memory pattern tends to stabilize the system in the 
  state of enhanced memory capacity.  We shall show that the 
  strength of the effect maximizes at the latest by the time the system 
  emits half of its energy.  At the same time, the information stored in the memory becomes accessible. Because of very transparent physical mechanism behind these findings, 
  it is almost obvious that it must be shared by generic systems 
  of enhanced memory capacity, including black holes   
  and  de Sitter Hubble patch.  This means that the tendency of a growing 
 back reaction from the memory burden must be applicable to such 
 objects. It is therefore expected that the back reaction  
 from a stored quantum information must drastically modify 
 the semi-classical picture by half-decay. 
 
 Note that this statement is in no conflict with any known 
 black hole property derived in semi-classical theory. 
 The reason, as already predicted both by calculations in
 gravitational microscopic theory  \cite{Dvali:2013eja,Dvali:2012rt,Dvali:2012wq,Dvali:2013vxa,Dvali:2017eba} as well as by 
 analysis of the prototype models \cite{1207.4059, 1507.02948,1507.08952, 
1601.01329, 1711.09079,1801.03918, 1712.02233, 1805.10292, 1810.02336, 1812.08749, 1906.03530, 1907.07332, 2003.05546} 
 including the present paper, is that after losing half of the  
 mass the semi-classical description is no longer applicable. 
 Namely:
 
{\it An old black hole that lost half of its mass is by no means
	equivalent to a young  classical black hole of the equal mass. } 
 
 In other words, the quantum effects such as the memory burden provide a universal {\it quantum clock }  
 that breaks the self-similarity of black hole evaporation 
 and suppresses its quantum decay. 
 This is the key result of the present paper.

 Some applications of this phenomenon to inflationary cosmology were already 
  discussed in \cite{1812.08749}. It was pointed out there that the
  memory burden of the primordial information carried  
  by degrees of freedom responsible for Gibbons-Hawking 
  entropy \cite{PhysRev.D15.2738} can provide a new type of the {\it cosmic  quantum hair}. This hair imprints a primordial quantum 
information from the early stages of inflation past last the $60$ 
e-foldings. 
 
 Now, the speculative part of our paper concerns the extrapolation 
 of the stabilization tendency for a black hole 
  beyond its half-decay.   At the present level 
  of understanding, such an extrapolation is a pure speculation since
 strictly speaking there is no guarantee that
  universality of the phenomenon holds on such long timescales. 
    In particular, it remains a viable option that after losing on the order of half 
   of its mass via quantum emission, a new classical instability can set in. 
   The black hole then can fall apart  via a highly non-linear 
   classical process.
  
 Our precautions can be explained in the following way.
  The state of maximal memory capacity represents a type of criticality. 
 The behavior of the system is then fully controlled by the gapless spectrum that 
  emerges in this state.
This explains the universal behavior of very different systems that 
 exist in such a state. 
  However, after half-decay the departure from the    
  critical state is significant and it is conceivable that different systems 
  behave differently after this point. So, despite the fact that 
  our prototype model gets stabilized, a real black hole
 could follow a different path.

In summary, the following two natural possibilities emerge: 
 \begin{itemize}
  \item The black hole continues its quantum decay but with an  
  extremely-suppressed emission rate. 
  \item A classical instability sets in and the black hole
  decays into highly non-linear graviton lumps, a sort of a 
  gravitational burst. 
\end{itemize} 
 
 While both outcomes would obviously have spectacular consequences,  
 in the present paper we shall speculate more on the first option.
 That is,  we can take
 the stabilization exhibited by the prototype model as a circumstantial 
 evidence indicating that large black holes  (i.e., $N_c \gg 1$, meaning much heavier than the Planck mass)
 behave in a similar manner, at least qualitatively.  
  Therefore, the decay rate of {\it macroscopic} black holes 
 could fall drastically after they lose half of their mass.  
  Not surprisingly, the consequences of such stabilization would be 
 dramatic.  One obvious application that shall be 
 discussed later is for primordial black holes as dark matter candidates.

\subsection{Outline}
 We shall now briefly outline our analysis in more 
 technical terms. 
The first ingredient is to explain the essence of the universal 
mechanism which allows the system to reach the state  
of enhanced memory storage. This mechanism is at work in  
large class of systems \cite{1207.4059, 1507.02948, 1507.08952, 1601.01329, 1711.09079,1801.03918,1712.02233,1805.10292, 1810.02336, 1812.08749}. 
Following \cite{1805.10292}, we shall refer to it as  \textit{assisted gaplessness}. 
 The essence of this mechanism  is most transparently explained by a simple model discussed in \cite{1711.09079,1801.03918,1712.02233, 1810.02336, 
1812.08749}, which we shall adopt as the prototype system in our analysis. 
The idea is that a high occupation of a particular low-frequency mode
to a certain critical level, 
renders a large set of other would-be-high-frequency modes gapless. 
 Namely, 
the highly-occupied \textit{master mode} interacts attractively with a set of other modes. We shall call the latter degrees of freedom the \textit{memory modes.}
Near the vacuum, the memory modes would have high energy gaps and would be useless (i.e., energetically costly) for storing information. 
However, coupling to the master mode lowers their energy gaps 
when the latter is highly occupied. That is, the attractive coupling with 
the master mode translates as a negative contribution to the energy of the memory modes. As soon as the occupation number of the master mode reaches a critical level $N_c$,
it can balance their positive kinetic energies. In this way, the memory modes become effectively gapless. Consequently, the states that correspond 
to different occupation numbers of these modes, 
$\ket{n_1,\ldots,n_K}$, are degenerate in energy
and contribute into the microstate entropy $S$.

So far, we have discussed how the master mode influences the memory modes 
by making them effectively gapless. However, the memory modes also backreact on the master mode via the effect of \textit{memory burden} \cite{1810.02336, 1812.08749}. Since the memory modes are gapless exclusively for a critical occupation $N_c$ of the master mode, any evolution of the latter away from such a state would cost a lot of energy. Therefore, the system backreacts on the master mode and 
resists to the change of its occupation number $N_c$. 

The next step in this line of research is to study to what extent the memory burden can be avoided. Namely, it has been proposed in \cite{1810.02336} that it can be alleviated if the system has a possibility 
of rewriting the stored information from one set of modes to another.  
This could happen if another set of memory modes exists, which becomes gapless at a different occupation number $N'_c$ of the master mode. Then, it is in principle conceivable that the 
occupation $N_c$
changes in time, provided this change is accompanied by 
rewriting of  the stored information from the first set of the memory 
modes to the second one. However, it has not yet been studied if such a rewriting is dynamically possible. 

In order to attack the issue, we can split this question in two parts.
\begin{enumerate}
	\item Does rewriting take place at all and under what conditions?
	\item If the answer to the first question is positive, what is the timescale of this process?
\end{enumerate}
Answering these questions is the goal of the present work.

In section \ref{sec:model}, we will summarize some of the findings of \cite{1207.4059, 1507.02948,1507.08952, 1601.01329,1711.09079,1801.03918,1712.02233,1805.10292, 1810.02336, 1812.08749} and develop a concrete prototype model that possesses states of enhanced memory capacity. Moreover, we 
discuss how it can be mapped on a black hole. In section \ref{sec:numerical}, we investigate the prototype model numerically and in particular study rewriting between two different sets of memory modes.
In short, we confirm that such a possibility can be realized dynamically.
 However, we discover that the speed of transition decreases as the size of the system increases. 
Applied to black holes, this indicates
that evaporation has to slow down at the latest after they have lost half 
of their mass. In section \ref{sec:NNC}, we briefly discuss systems that do not conserve particle number and point out that our conclusions also apply to them.  Section \ref{sec:primoridalBlackHoles} is dedicated to studying consequences of our findings for primordial black holes as dark matter candidates.
We give an outlook in section \ref{sec:outlook} and the appendix contains 
details as to how the speed of rewriting depends on the various parameters of the system.

\section{Enhanced Memory Storage: A Prototype Model}
\label{sec:model}
\subsection{Assisted Gaplessness}
Following \cite{1712.02233, 1801.03918, 1805.10292}, we shall construct a 
simple prototype model that dynamically achieves a state 
with many gapless modes and correspondingly a high microstate entropy $S$. We consider $K$ bosonic modes, which we describe by the usual creation and annihilation operators $\hat{a}_k^\dagger$, $\hat{a}_k$, where $k=1,\ldots K$. They satisfy standard commutation relations (here and throughout $\hbar =1$):
 \begin{equation} 
[\hat{a}_j,\hat{a}_k^{\dagger}] = \delta_{jk}\,,\qquad
[\hat{a}_j,\hat{a}_k]  =   [\hat{a}_j^{\dagger},\hat{a}_k^{\dagger}] =0\,,  
\label{algebra} 
\end{equation} 
and the corresponding number operators are given as $\hat{n}_k=\hat{a}_k^\dagger \hat{a}_k$. We denote its eigenstates by $\ket{n_k}$, where $n_k$ is the eigenvalue. Moreover, we label the energy gap of $\hat{n}_k$ by 
$\epsilon_k$.

Using the modes $\hat{n}_k$, one can form the states
\begin{equation} \label{memoryStates}
	\ket{n_1,\ldots,n_K} \equiv  \ket{n_1}\otimes \ket{n_2}\otimes,\ldots,\otimes \ket{n_K} \,,
\end{equation}
where $n_1,\ldots n_K$ can take arbitrary values. Clearly, the number of such states scales exponentially with $K$, so for $K\gtrsim S$, their number reaches the required number of microstates. However, it is important to consider the energy of the states \eqref{memoryStates}. Namely, two different states $\ket{n_1, \ldots n_S}$ and $\ket{n'_1, \ldots n'_S}$ differ by the energy $\Delta E = \sum_{k=1}^K \epsilon_k (n_k-n'_k)$. So unless the $\epsilon_k$ are extremely small, the states corresponding to different occupation numbers of the modes $\hat{n}_k$ are not degenerate in energy. Therefore, they cannot contribute to a microstate entropy.

We can change this situation by introducing another mode $\hat{n}_0$, with creation and annihilation operators $\hat{a}_0^\dagger$, $\hat{a}_0$ and commutation relations analogous to \Eq \eqref{algebra}. The key point is that we add an attractive interaction between the mode $\hat{n}_0$ and all other modes:
\begin{equation}
	\hat{H} = \epsilon_0 \hat{n}_0 + \left(1-\frac{\hat{n}_0}{N_c}\right) \sum_{k =1}^K \epsilon_k \hat{n}_k \,,
\end{equation}
where we parameterize the strength of the interaction by $1/N_c$ with $N_c\gg 1$. As long as $\hat{n}_0$ is not occupied, the gaps of the $\hat{n}_k$-modes are still given by $\epsilon_k$. As soon as we populate $\hat{n}_0$, however, the effective gaps $\mathcal{E}_k$ of the $\hat{n}_k$-modes are lowered:
\begin{equation} \label{effectiveGap}
	 \mathcal{E}_k = \left(1-\frac{n_0}{N_c} \right)\epsilon_k \,.
\end{equation}
Once a critical occupation $n_0 = N_c$ is reached, all modes $\hat{n}_k$ become effectively gapless, 	$\mathcal{E}_k=0$. 

Therefore, all states of the form
\begin{equation}
	\ket{\underbrace{N_c}_{n_0}, n_1, \ldots, n_K}
	\label{Mpattern}
\end{equation}
are degenerate in energy for arbitrary values of $n_1$, \ldots, $n_K$. In 
this situation, $\hat{n}_0$ is the master mode, which assists the memory modes $\hat{n}_1$, \ldots, $\hat{n}_K$ in becoming gapless. We note, however, that one has to invest the energy $\epsilon_0 N_c$ to achieve gaplessness. If each of the memory modes can have a maximal occupation of $d$, this leads to a number of $(d+1)^K$ distinct states that possess the same 
energy, \ie an entropy
\begin{equation}
	S = K \ln (d+1) \,.
\end{equation}
In this way, a large number $K\approx S$ of nearly-gapless modes leads to 
the entropy $S$. 

\subsection{Memory Burden}
\label{ssec:memoryBurden}

Following \cite{1810.02336, 1812.08749}, we next investigate the effect of memory burden, i.e., how the memory modes backreact on the master mode. 
To this end, we add another mode with which $\hat{n}_0$ can exchange occupation number. We denote its creation and annihilation operators by $\hat{b}_0^\dagger$, $\hat{b}_0$ with commutation relations analogous to \Eq \eqref{algebra} and $\hat{m}_0 = \hat{b}_0^\dagger \hat{b}_0$ is the corresponding number operator. Then the Hamiltonian becomes 
\begin{equation} \label{HamiltonianBandA}
\hat{H} = \epsilon_0 \hat{n}_0 + \epsilon_0 \hat{m}_0 +  \left(1-\frac{\hat{n}_0}{N_c}\right) \sum_{k =1}^K \epsilon_k \hat{n}_k +  C_0\left( \hat{a}_0^\dagger \hat{b}_0 + \hat{b}_0^\dagger \hat{a}_0\right)   \,,
\end{equation}
where $C_0$ parametrizes the strength of interaction between $\hat{n}_0$ and $\hat{m}_0$. We choose the gap of $\hat{n}_0$ and $\hat{m}_0$ to be equal in order to facilitate the oscillations between them.

Now we consider the initial state
\begin{equation}
\ket{\text{in}_1} = \ket{\underbrace{N_c}_{n_0}, \underbrace{0}_{m_0}, n_1, \ldots, n_K} \,.
\end{equation} 
Since the occupation number of each of the memory modes is conserved in time, it is possible to solve the system analytically. For the expectation 
value of $\hat{n}_0$, one obtains \cite{1810.02336}: 
\begin{equation} \label{expectationValueAnalytic}
n_0(t) = N_c\left(1- \frac{4C_0^2}{4C_0^2 + \mu^2} \sin^2(\sqrt{C_0^2 + 
\mu^2/4} \, t)\right) \,,
\end{equation} 
where we defined
\begin{equation} \label{memoryBurden}
\mu = -\sum_{k =1}^K \epsilon_k n_k/N_c \,.
\end{equation}
This quantity characterizes the strength of memory burden, as we shall demonstrate. It is related to the effective energy gaps \eqref{effectiveGap} as 
\begin{equation}
	\mu = \sum_{k =1}^K n_k \frac{\partial  \mathcal{E}_k}{\partial n_0} 
\,.
\end{equation}

\begin{figure*}
	\begin{subfigure}{0.3\textwidth}
		\includegraphics[width=\textwidth]{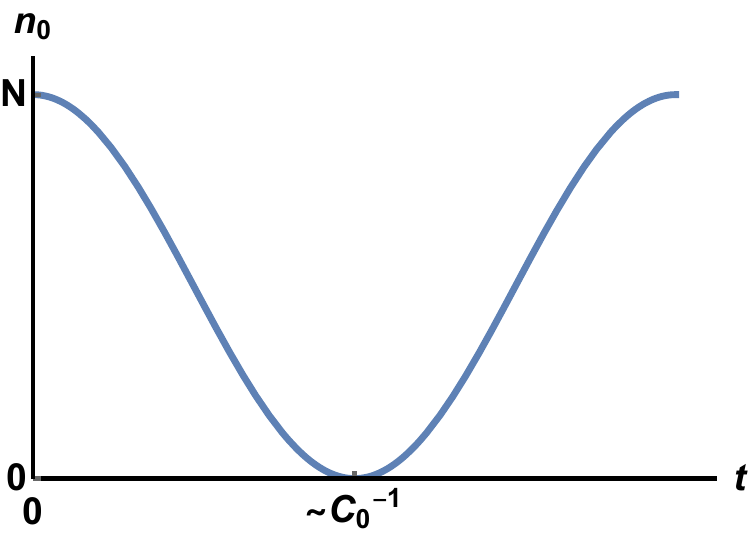}
		\caption{$\mu=0$: absence of memory burden. The expectation value $n_0$ oscillates freely with full amplitude.} 
		\label{sfig:noBurden}
	\end{subfigure}
	\begin{subfigure}{0.3\textwidth}
		\includegraphics[width=\textwidth]{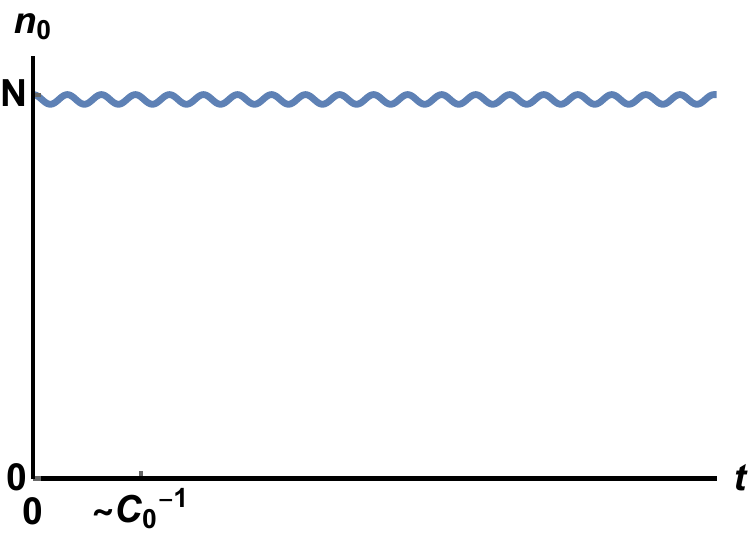}
		\caption{$|\mu|\ =2.5$ (and $p=1$): early backreaction due to memory burden ties $n_0$ to its initial value.}
		\label{sfig:fullBurden}
	\end{subfigure}
	\begin{subfigure}{0.3\textwidth}
		\includegraphics[width=\textwidth]{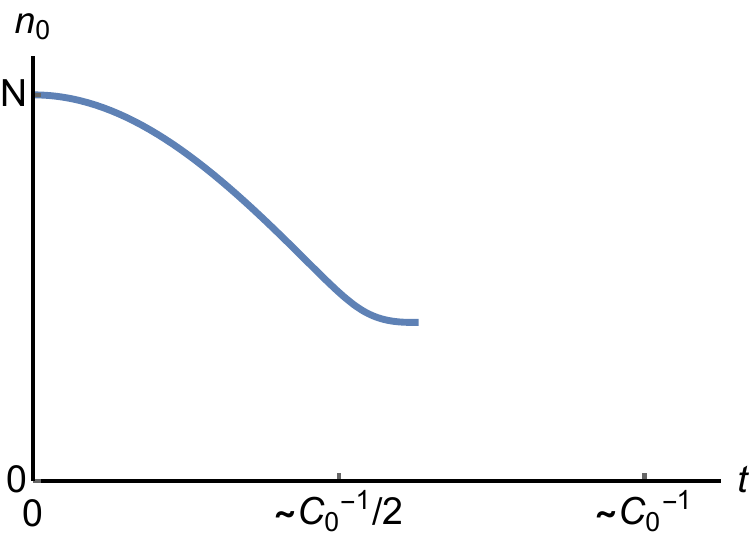}
		\caption{$|\mu|\ =2.5$ and $p=8$: memory burden can be delayed up to a timescale on the order of half decay. Backreaction sets in at the latest at that point and stabilizes the system.}
		\label{sfig:delayedBurden}
	\end{subfigure}
	\caption{Plots of the time evolution of $n_0$ for $N_c=25$ and $C_0 = 
\epsilon_0/\sqrt{N_c} = 1/5$. \figs \ref{sfig:noBurden} and \ref{sfig:fullBurden} follow from \eqref{expectationValueAnalytic}. \fig \ref{sfig:delayedBurden} is an approximate solution of the system \eqref{HamiltonianHigherOrder}.}
	\label{fig:burden}
\end{figure*}

\Eq \eqref{expectationValueAnalytic} shows that the memory modes drastically influence the time evolution of $\hat{n}_0$. First, we consider the special situation in which all memory modes are unoccupied. This implies $\mu = 0$ so that $n_0(t)$ performs oscillation with maximal amplitude on a timescale of $C_0^{-1}$. This behavior, which is depicted in \fig \ref{sfig:noBurden}, is identical to the case in which the memory modes do not exist. For $n_k \neq 0$, this situation changes as soon as  either the 
occupation of the memory modes is high enough or their free gaps $\epsilon_k$ are sufficiently big. In both cases, one gets $C_0^2/\mu^2 \ll 1$ so 
that the amplitude of oscillations is suppressed by this ratio. This is shown in \fig \ref{sfig:fullBurden} for exemplary values of the parameters. Thus, the stored information ties $\hat{n}_0$ to its initial state. This is the essence of memory burden.

Consequently, the crucial question arises to what extent memory burden can be avoided. Following \cite{1812.08749}, a first way consists in modifying the Hamiltonian \eqref{HamiltonianBandA} as follows:
\begin{equation} \label{HamiltonianHigherOrder}
\hat{H} = \epsilon_0 \hat{n}_0 + \epsilon_0 \hat{m}_0 +  \left(1-\frac{\hat{n}_0}{N_c}\right)^p \sum_{k =1}^K \epsilon_k \hat{n}_k +  C_0\left( \hat{a}_0^\dagger \hat{b}_0 + \hat{b}_0^\dagger \hat{a}_0\right)   \,.
\end{equation}
 In this case, the effective energy gaps read
\begin{equation} \label{effectiveGapHigherOrder}
\tilde{\mathcal{E}}_k = \left(1-\frac{n_0}{N_c} \right)^p\epsilon_k \,.
\end{equation}
Consequently, the memory burden becomes
\begin{equation} \label{smallerBurden}
\tilde{\mu} =p \left(\frac{N_c-n_0}{N_c}\right)^{p-1}\mu \,,
\end{equation}
where we expressed it in terms of \Eq \eqref{memoryBurden}. We see that it gets suppressed by powers of $(N_c-n_0)/N_c$. The larger $p$ is, the more the backreaction gets delayed. However, it sets in at the latest when $\tilde{\mu}$ assumes the critical value $C_0$:
\begin{equation}
N_c-n_0 = N_c \left(\frac{C_0}{p |\mu|}\right)^{1/(p-1)}\,.
\end{equation}
For $\mu > C_0$, it is clear that memory burden can no longer be avoided as soon as $N_c-n_0 $ is of the order of $N_c$. Thus, backreaction becomes important and the system stabilizes at the latest after a timescale on the order of half decay, as is exemplified in \fig \ref{sfig:delayedBurden}.

\subsection{Avoiding Memory Burden by Rewriting}\label{ssec:avoidingMemoryBurden}
In \cite{1810.02336}, another way of alleviating memory burden was proposed. The idea is to introduce a second sector of memory modes, which are not occupied in the beginning, but which can exchange occupation number with the first sector. If the coupling of the second sector to the master mode is such that it becomes gapless for a smaller value $n_0 = N_c-\Delta N_c$, then a final state in which $n_0$ has diminished by $\Delta N_c$ 
and all excitations have been transferred from the first to the second memory sector becomes energetically available.

We denote the creation and annihilation operators of the second memory sector by $\hat{a}^{'\dagger}_{k'}$, $\hat{a}^{'}_{k'}$, the number operator by $\hat{n}'_{k'} =\hat{a}^{'\dagger}_{k'} \hat{a}^{'}_{k'}$ and assume  the usual commutation relations \eqref{algebra}, where $k'=1,\ldots, K'$. Then the Hamiltonian becomes:
\begin{align}
\hat{H} &= \epsilon_0 \hat{n}_0 + \epsilon_0 \hat{m}_0  +  C_0\left( \hat{a}_0^\dagger \hat{b}_0 + \hat{b}_0^\dagger \hat{a}_0\right) 
\nonumber\\
& +  \left(1-\frac{\hat{n}_0}{N_c}\right) \sum_{k =1}^K \epsilon_k \hat{n}_k
+  \left(1-\frac{\hat{n}_0}{N_c-\Delta N_c}\right) \sum_{k'=1}^{K'} \epsilon_{k'} \hat{n}'_{k'} 
\nonumber\\
& + \sum_{k=1}^{K}\sum_{k'=1}^{K'} C_{k,k'}\left( \hat{a}_k^\dagger \hat{a}'_{k'} +  \text{h.c.}\right) 
\nonumber \\
& +\sum_{k=1}^{K}\sum_{\substack{l=1\\l> k}}^{K} \tilde{C}_{k,l}\left( \hat{a}_k^\dagger \hat{a}_l +  \text{h.c.}\right)
\nonumber\\
& + \sum_{k'=1}^{K'}\sum_{\substack{l'=1\\l'> k'}}^{K'}  \tilde{C}_{k',l'}\left( \hat{a}_{k'}^{'\dagger} \hat{a}'_{l'} +  \text{h.c.}\right) \,.
\label{fullHamiltonian}
\end{align}

Here the parameters $C_{k,k'}$ determine the strength of coupling between 
the two memory sectors. For completeness, we have moreover introduced interactions within each memory sector, the strength of which is set by $ \tilde{C}_{k,l}$. In order to maximize the effect of memory burden, we have 
set $p=1$.

We consider an initial state in which the first memory sector is gapless and only the first sector is occupied:
\begin{equation} \label{initialState}
\ket{\text{in}}=\ket{\underbrace{N_c}_{n_0},\underbrace{0}_{m_0},n_1, \ldots n_K, \underbrace{0}_{n'_1}, \ldots, \underbrace{0}_{n'_{K'}}} \,.
\end{equation}
As explained, a final state is energetically available in which the second memory sector is gapless and only the second sector is occupied:
\begin{equation} \label{finalState}
\ket{\text{out}}=\ket{\underbrace{N_c-\Delta N_c}_{n_0},\underbrace{\Delta N_c}_{m_0},\underbrace{0}_{n_1}, \ldots \underbrace{0}_{n_K}, n'_1, \ldots, n'_{K'}} \,.
\end{equation}
The total occupation in the two memory sectors, 
\begin{equation}
N_m \equiv	\sum_{k =1}^K n_k + \sum_{k' =1}^{K'} n'_{k'} \,, 
\end{equation}
is conserved.

Within our setup, we require that near the initial state \eqref{initialState}, the second memory sector is not gapless. The mildest possible constraint that realizes this is (see also \Eq \eqref{epsilonEffCondition} shortly below)
\begin{equation} \label{boundEpsilonPrime}
\left|{\mathcal E}^{'}_k\right|\gg \frac{\epsilon_0}{\sqrt{N_m}} \,.
\end{equation}
Alternatively, we can also impose the stronger bound
\begin{equation} \label{boundEpsilonPrimeStronger}
	\left|{\mathcal E}^{'}_k\right|\gg \epsilon_0 \,.
\end{equation}
Throughout, we assume that at least the milder constraint \eqref{boundEpsilonPrime} is fulfilled.

Once the state \eqref{finalState} exists in the spectrum, it is no longer 
{\it a priori} excluded that the system evolves away from the initial state. So in principle, it becomes possible to avoid memory burden by simultaneously rewriting information from the $\hat{a}^{\dagger}_{k}$, $\hat{a}_{k}$-modes to the $\hat{a}^{'\dagger}_{k'}$, $\hat{a}^{'}_{k'}$-ones. However, by no means does this imply that the system will dynamically evolve from $\ket{\text{in}}$ to $\ket{\text{out}}$ on a reasonable time scale. Therefore, we shall study if and under what conditions this transition actually takes place.

\subsection{Bounds on Couplings}
Before we investigate the time evolution, we study how large the couplings of the memory modes can be. Namely, they must fulfill the condition that the effective gap $\mathcal{E}_{\text{eff}}$ of the memory modes stays close to zero in the presence of couplings. In order to obtain the mildest possible bound, we can consider a situation in which the gaps can equally be offset to positive or negative values. Consequently, occupying $N_m$ modes typically only gives an energy disturbance of $\sqrt{N_m} \mathcal{E}_{\text{eff}}$. Imposing that it is smaller than the elementary gap, we obtain the constraint\footnote{\label{footnote:random}Note that without assuming contributions with random signs the constraint is $\mathcal{E}_{\text{eff}} \lesssim \frac{\epsilon_0}{N_m}$.
}
\begin{equation} \label{epsilonEffCondition}
\mathcal{E}_{\text{eff}} \lesssim \frac{\epsilon_0}{\sqrt{N_m}} \,.
\end{equation}

First, we will turn to the coupling $\tilde{C}_{k,l}$ within one memory sector, where we assume that all $\tilde{C}_{k,l}$ are of the same order. If we only consider two modes for a moment, they are described by the effective coupling matrix
\begin{equation} \label{gaplessCouplingMatrix}
\begin{pmatrix}
0 & \tilde{C}_{k,l}\\
\tilde{C}_{k,l} &0 
\end{pmatrix}\,.
\end{equation}
Thus, disturbing the gap by at most $\epsilon_0/\sqrt{N_m}$ implies $\tilde{C}_{k,l} \lesssim \epsilon_0/\sqrt{N_m}$. However, we need to take into account that it couples to many modes.  When we view the couplings within one memory sector as samples from identical independent distributions with zero mean and unit variance, then the corresponding matrix, \ie the generalization of \Eq \eqref{gaplessCouplingMatrix} to many modes, belongs to a Wigner Hermitian matrix ensemble. In this situation, Wigner's semicircle law states that the spectral distribution converges, and in particular becomes independent of the dimension $K$, if the entries of the matrix are rescaled by $1/\sqrt{K}$ (see \eg \cite{RMT}). Thus, we need to suppress the coupling constants with this factor to maintain  approximate gaplessness for the majority of modes:
\begin{equation} \label{boundCm}
\tilde{C}_{k,l} \lesssim \frac{\epsilon_0}{\sqrt{N_m} \sqrt{K}} \,.
\end{equation}

We can also arrive at the same conclusion by studying the expectation value of the off-diagonal elements in the Hamiltonian, as was done in \cite{1804.06154}. This energy scales as $N_m \tilde{C}_{k,l}$, where we took into account that $N_m^2$ non-zero entries only give a contribution on the 
order of $N_m$ as long as there are both positive and negative summands. Requiring this energy to be smaller than $\epsilon_0$ yields
\begin{equation} \label{boundCm2}
\tilde{C}_{k,l} \lesssim \frac{\epsilon_0}{N_m} \,.
\end{equation}
For a typical occupation $N_m \sim K$, this bound is identical to \Eq \eqref{boundCm}

Next, we study the coupling $C_{k,k'}$ of modes from different memory sectors, where we assume again that all $C_{k,k'}$ are of the same order. In 
this case, we get the effective coupling matrix 
\begin{equation} \label{gapCouplingMatrix}
\begin{pmatrix}
0 & C_{k,k'}\\
C_{k,k'} & \epsilon_k\Delta N_c/N_c 
\end{pmatrix}\,.
\end{equation}
We estimate $\mathcal{E}_{\text{eff}}  \sim K C_{k,k'}^2N_c/(\epsilon_k\Delta N_c)$. This gives the constraint
\begin{equation}
C_{k,k'} \lesssim \frac{\sqrt{\epsilon_0 \epsilon_k \Delta N_c}}{\sqrt{K N_c} (N_m)^{1/4}} \,,
\end{equation}
which is milder than the bound \eqref{boundCm} since $\epsilon_k\geq\epsilon_0$.

\subsection{Application to Black Holes}
\label{ssec:blackHole}

 We now wish to apply our results to black holes. 
Naturally, we shall work under the assumption 
that the above quantum system of enhanced memory capacity 
captures some very general properties 
of black hole information storage.  Ideally, we would like not to be
confined to any particular microscopic theory but rather 
to make use of certain universal properties that any such theory must 
incorporate. For example, existence of modes
that become gapless around a macrostate 
corresponding to a black hole is expected to be such 
universal property.  Indeed, without gapless excitations, it would be 
impossible to account for the black hole microstate entropy. 
Also, an important fact is that  
the Bekenstein-Hawking entropy  \cite{PhysRev.D7.2333}  
depends on the black hole mass $M$, 
 \begin{equation} \label{SBH}
	S = 4\pi G_NM^2\,, 
\end{equation} 
where $G_N$ is Newton's constant. 
Correspondingly, the number of the gapless modes 
that a black hole supports depends on its mass. 
This makes it obvious that any evolution that deceases the black hole 
mass must affect the energy gaps of the memory modes. That is, when 
a black hole evaporates, some of the modes that were previously gapless now must acquire the energy gaps.  
But then, this process must result into a memory burden effect that resists against the decrease of  $M$.  This is the main lesson that we learn about black holes from our 
analysis.  For more quantitative understanding, we shall try to 
choose the parameters of the above toy model to be maximally close 
to the corresponding black hole characteristics.  

For a crude guideline, it will be useful to keep in mind a particular 
microscopic theory of black hole quantum $N$-portrait \cite{1112.3359}. 
Although we wish to keep our analysis maximally general, 
having a microscopic theory helps in establishing a precise dictionary  
between parameters of a black hole and the presented simple 
Hamiltonian.  It also shows how well the seemingly toy model 
captures the essence of the phenomenon.

 According to quantum $N$-portrait,  a black hole of 
Schwarzschild radius $r_g =2 G_N M$,   represents a saturated bound-state of soft gravitons. The 
characteristic wavelength of gravitons contributing into the gravitational self-energy is given by  $r_g$.  These constituent gravitons play the role of
the master mode. Namely, their occupation number is critical and this 
renders a set of other modes gapless. The latter modes play the role
of the memory modes. Without the presence of a critical occupation number 
of the master mode, the memory modes would represent free gravitons 
 of very high frequencies and respectively would possess very high energy 
gaps.
 That is, it would be very costly in energy to excite those modes 
if the  occupation number of the master mode were not critical.  
   Now, the idea is that these gapless  modes account for the Bekenstein-Hawking entropy (\ref{SBH}).  
In the above toy model, the role of the master mode is 
played by $\hat{a}_0$ with occupation number $n_0$.  

In this picture, the Hawking radiation \cite{Hawking} is a result of quantum depletion. Consequently, some of the particles of the master mode get 
converted into free quanta and the occupation number of the master mode decreases.  
 These free Hawking quanta are impersonated by 
the quanta  of mode $\hat{b}_0$ and they have the occupation number $m_0$.  Initially, $m_0=0$.  However, during the conversion the occupation 
number $m_0$ increases while 
$n_0$ decreases and moves away from the critical value. 
This is expected to create a memory burden effect. Of course, unlike a black hole, the model \eqref{fullHamiltonian} performs oscillations, i.e., $\hat{m}_0$ again loses quanta after a certain timescale. Thus, we can map our model on a black hole only up to this point, but this fully suffices for our conclusions\footnote
{The bilinear coupling between modes is motivated as the simplest possible coupling that is able to effectively describe energy transfer between degrees of freedom.
In order to model a decay more precisely, one could instead consider a coupling to many species, $\frac{C_0}{\sqrt{F}}\sum_{f=1}^{F}\hat{a}_0 \hat{b}^{\dagger}_f + \textit{h.c.}$ (all with the same gap $\epsilon_f = 
\epsilon_0$), which could e.g., represent momentum modes of a field-theoretic system. 
In the limit of large $F$, one can achieve strict decay with the same rate as in \eqref{fullHamiltonian}.
}.
Another reason why the timescale of validity of our model is limited is that black holes can exist for all values of the mass $M$. Therefore, a tower of sets of momentum modes has to exist so that one of them becomes gapless for each value of $M$. In contrast, we only consider two sets of momentum modes in our model. For this reason, our model can no longer be mapped on a black hole as soon as a third set of momentum modes would start 
to be populated.
Finally, particle number in gravity is not conserved, unlike in our prototype model. We shall discuss in section \ref{sec:NNC} why this does not change our conclusions. 

We can now choose the parameters in such a way that Hamiltonian \eqref{fullHamiltonian} reproduces the generic information-theoretic properties of 
a black hole. First, we set the elementary gap as $\epsilon_0=r_g^{-1}$ 
to make sure that Hawking quanta have the correct typical energy $r_g^{-1}$. Next, we need $K=S$ to obtain the desired entropy. Consequently, a typical pattern has $N_m=S/2$, since for large black holes $S\gg 1$ and 
the number of patterns with different $N_m$ is insignificant.
We can also estimate the gap of the memory modes. 
Since the system is spherically symmetric, we can label states by the quantum numbers $(l,m)$ of angular harmonics. Assuming no significant part of the energy of the modes is in radial motion, we need to occupy states at least until $l\sim \sqrt{K}$ in order to obtain a number of $K$ modes, since the degeneracy of each level scales as $l$. In this case, the highest mode has an energy of $\epsilon_k=\sqrt{K} \epsilon_0$. We use this scale to estimate the free gap of the memory modes because the relative split in energy among the levels is inessential for our discussion. We remark that this means that those modes are Planckian, $\epsilon_k \sim 1/\sqrt{G_N}$. Finally, we have freedom in choosing the critical occupation number $N_c$. For concreteness, we set $N_c=S$, as is motivated by the quantum N-portrait.
In this way, the total energy of the system reproduces the mass of the black hole: $M = N_c \epsilon_0$. 
In summary, we can express all quantities in terms of the entropy and the 
Schwarzschild radius:
\begin{align} \label{parametersBH}
\epsilon_0 &=r_g^{-1} \,, \qquad N_c=S\,, \qquad K=S\,,\qquad N_m=S/2\,,
\nonumber \\ 
\epsilon_k &= \sqrt{S} r_g^{-1} \,.
\end{align}

Since gravitational coupling is universal, all $C_{k,k'}$ and all $\tilde{C}_{k,l}$ need to be of the same order. So \Eq \eqref{boundCm} gives the 
strongest constraint, which reads
\begin{equation} \label{boundCmBH}
C_{k,k'} \sim \tilde{C}_{k,l} \lesssim \frac{\epsilon_0}{S} \,.
\end{equation}
As stated before, this bound is the softest possible one. In real black holes constraints may be stronger.

When applying our analysis to real black holes, 
some additional facts must be taken into account.
Namely, together with the $\hat{b}_0$-mode, which impersonates the outgoing 
free quanta of Hawking radiation, there are also the free modes of higher 
momenta. In particular, there will of course exist  
the free modes of the same momenta $k$ as the memory modes  
$\hat{a}_k$.  These modes are denoted by  $\hat{b}_k$. 
Now, unlike the memory $\hat{a}_k$-modes, the  $\hat{b}_k$-modes are 
{\it not} subjected to the assisted gaplessness. Correspondingly,  they satisfy the dispersion relations of free quanta.  
 That is, the frequencies $\epsilon_k$ of 
 $\hat{b}_k$-modes are of order of their momenta and, therefore, are much 
higher than the frequencies of the corresponding 
 $\hat{a}_k$-modes.  The essence of the situation is described by the following Hamiltonian  
   
\begin{align}
\hat{H}_{\text{higher}} &=\sum_{k=1}^{K} \epsilon_k \hat{b}_k^\dagger 
\hat{b}_k +  \sum_{k=1}^{K} C_k \left(\hat{a}_k^\dagger \hat{b}_k + \hat{b}_k^\dagger \hat{a}_k\right) 
\nonumber \\
&+  \sum_{k'=1}^{K'} C_{k'} \left(\hat{a}_{k'}^{'\dagger} \hat{b}_{k'} + \hat{b}_{k'}^\dagger \hat{a}'_{k'}\right) \,.
\end{align}
As before, we have $\epsilon_k=\sqrt{S}\epsilon_0$.

Now, the values of the couplings $C_k$ can be deduced from the 
consistency requirement that they do not disturb the gaplessness of the $\hat{a}_k$-modes. The corresponding coupling matrix is (for $n_0=N_c$) 
\begin{equation}
\begin{pmatrix}
0 & C_k\\
C_k & \epsilon_0\sqrt{S} 
\end{pmatrix}\,.
\end{equation}
From the condition that the vanishing gap is offset by at most $\epsilon_0/\sqrt{S}$, it follows that $C_k^2/(\epsilon_0\sqrt{S}) \lesssim \epsilon_0/\sqrt{S}$, \ie $C_k \lesssim \epsilon_0$.
Thus, due to enormous level splitting,  the mixing between the 
$\hat{a}_k$ and  $\hat{b}_k$ is highly suppressed. 
Correspondingly, the free modes ($\hat{b}_k$)
 of the same momenta as the memory 
modes ($\hat{a}_k$) stay unoccupied during the time evolution.  
In other words, the information encoded in the memory modes 
cannot be transferred to the outgoing radiation since the mixing 
between the two sets of modes is highly suppressed. 

This finding has important implications as it explains 
{\it microscopically} \cite{1810.02336}  why a black hole at the earliest 
stages of its evolution 
releases energy but almost no information. 
This fact is often considered as one of the mysteries 
of black hole physics. 
What we are observing is that this is a universal property 
shared by systems that are in a state of enhanced memory capacity due to
assisted gaplessness. The ``secret'' lies in a large level splitting between 
the memory modes subjected to the assisted gaplessness and 
their free counterparts. 

Since due to the above reason the $\hat{b}_k$-modes will largely stay unoccupied, we do not include them in the numerical simulations. 
Finally,  Since the gravitational interaction scales with energy, 
 we get a bound on the coupling $C_0$:
\begin{equation} \label{boundC0}
C_0 \lesssim \frac{\epsilon_0}{\sqrt{S}} \,.
\end{equation}

\section{Numerical Time Evolution}
\label{sec:numerical}
For the numerical study, we need to specialize to a particular realization of the system \eqref{fullHamiltonian}. In doing so, we keep in mind the 
special case of black holes, although our choice of parameters stays much 
more general. First, we choose the free gaps of all memory modes in both sectors to be equal, $\epsilon_k=\epsilon_{k'} =: \epsilon_m$. Moreover, we assume that all couplings $C_{k,k'}$ and $\tilde{C}_{k,l}$ are of the same order. Therefore, we can represent them as 
$C_{\text{m}} f_i(k,k')$, where $f_i(k,l)$ take values of order one. It is important that the $f_i(k,l)$ are non-trivial to break the exchange symmetry $\hat{a}_k \leftrightarrow \hat{a}_l$. We choose them so that they essentially take random values in $|f_i(k, l)|\in [0.5;1]$, with both plus- and minus-sign.\footnote
{Concretely, we choose $f_i(k,l) = 
\left\lbrace \begin{array}{rcl}
 F_i(k,l) -1 & \mbox{for} & F_i < 0.5
\\ 
F_i(k,l) & \mbox{for} & F_i\geq 0.5 
\end{array}\right.$, where $F_i(k,l) = \left( \sqrt{2}(k+\Delta k_i)^3 + \sqrt{7}(l+\Delta l_i)^5 \right) \mod 1$. Moreover, we set $\Delta k_1 = \Delta k_2 = 1$, $\Delta k_3 = K+1$, $\Delta l_1 = \Delta l_3 = 
K+1$ as well as $\Delta l_2 = 1$.}
Finally, we note that $\epsilon_0\left(\hat{n}_0 + \hat{m}_0\right)$ corresponds to a conserved quantity. Since as initial states we only consider 
eigenstates of this operator, it only leads to a trivial global phase and 
we can leave it out. In turn, we will use $\epsilon_0$ as basic energy unit.  We arrive at the Hamiltonian
\begin{align}
\frac{\hat{H}}{\epsilon_0} =  \frac{C_0}{\epsilon_0}& \left(  \hat{a}_0^\dagger \hat{b}_0 + \hat{b}_0^\dagger \hat{a}_0\right) 
+  \frac{\epsilon_m}{\epsilon_0}\left(1-\frac{\hat{n}_0}{N_c}\right) \sum_{k =1}^K \hat{n}_k 
\nonumber\\ + 
 \frac{\epsilon_m}{\epsilon_0} &\left(  1-\frac{\hat{n}_0}{N_c-\Delta N_c}\right) \sum_{k'=1}^{K'}  \hat{n}'_{k'} 
\nonumber\\ + 
   \frac{C_{\text{m}}}{\epsilon_0} &\Bigg\{ \sum_{k=1}^{K}\sum_{k'=1}^{K'} f_1(k,k')\left( \hat{a}_k^\dagger \hat{a}'_{k'} +  \text{h.c.}\right) 
 \nonumber\\
&+ \sum_{k=1}^{K}\sum_{\substack{l=1\\l> k}}^{K} f_2(k,l)\left( \hat{a}_k^\dagger \hat{a}_l +  \text{h.c.}\right) 
\nonumber\\ 
 &+\sum_{k'=1}^{K'}\sum_{\substack{l'=1\\l'> k'}}^{K'}  f_3(k',l')\left( \hat{a}_{k'}^{'\dagger} \hat{a}'_{l'} +  \text{h.c.}\right)\Bigg\} \,,
\label{fullHamiltonianSimple}
\end{align}
where we set $\epsilon_0=1$ from here on.

As a final simplification for the numerical study, we truncate all memory 
modes to qubits. Correspondingly, we consider the initial state
\begin{equation} \label{initialStateSimple}
	\ket{\text{in}}=\ket{N_c,0,\underbrace{1,\ldots, 1}_{N_m}, 0, \ldots, 0} \,,
\end{equation}
\ie $\hat{n}_0$ is populated with $N_c$ particles, $\hat{m}_0$ is empty and there is one particle in each of the first $N_m$ memory modes.

Unless otherwise stated, the values for the parameters we use are
\begin{align} \label{systemParameters}
	\epsilon_m &= \sqrt{20} \,, \quad N_c=20\,, \quad  \Delta N_c = 12 
\,, \quad K=K'=4 \,, 
\nonumber \\	
	 C_0 &= 0.01\,,\quad N_m=2 \,.
\end{align}
These parameters define both the Hamiltonian and the initial state, up to 
a choice of the coupling $C_{\text{m}}$.  We note that we chose $N_m = K/2$ since this corresponds to the most probable state in the limit of large $K$.\footnote{Since $K$ will be mapped onto the entropy of a BH, this 
reasoning only applies to macroscopic BHs with $M \gg M_p$.}

For the numerical time evolution, we use the approach and software developed in \cite{timeEvolver}. It is based on a Krylov subspace method and has the strength that it provides a rigorous upper bound on the numerical error, i.e., the norm of the difference between the exact time-evolved state and its numerical approximation. Throughout we set it to be $10^{-6}$, 
with the exception of systems with $K =8$, for which we use $10^{-5}$.

\subsection{Possibility of Rewriting}

\begin{figure*}
	\begin{subfigure}{0.4\textwidth}
		\includegraphics[width=\textwidth]{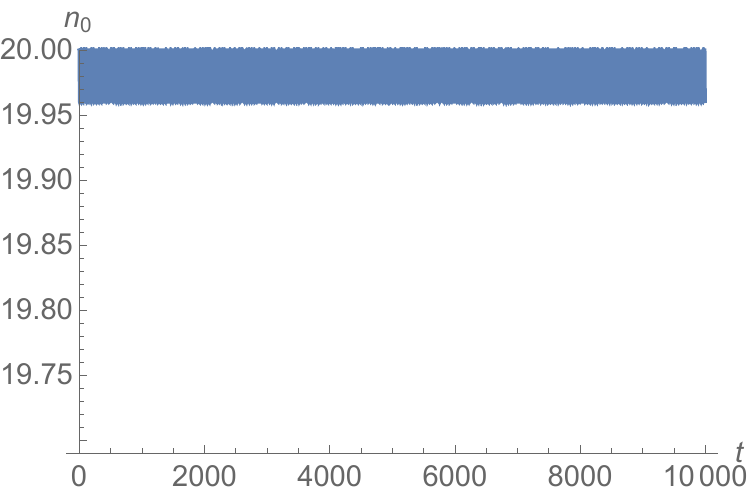}
		\hspace{0.18em} 
		\includegraphics[width=\textwidth]{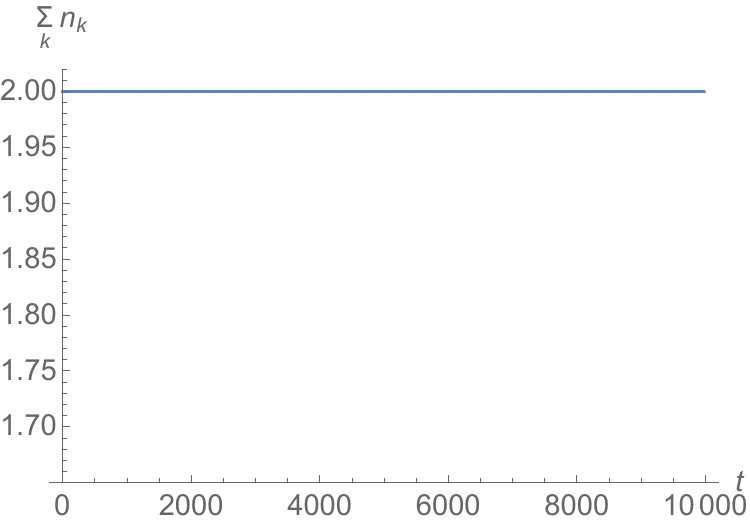}
		\caption{$C_{\text{m}} = 0$.} 
			\label{sfig:freeEvolution}
	\end{subfigure}
	\hspace{0.05\textwidth}
	\vspace{40pt}
	\begin{subfigure}{0.4\textwidth}
		\includegraphics[width=\textwidth]{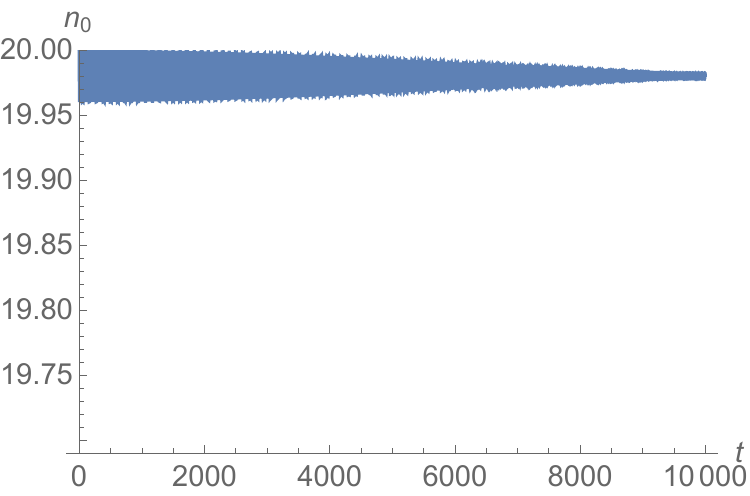}	
		\hspace{0.18em} 
		\includegraphics[width=\textwidth]{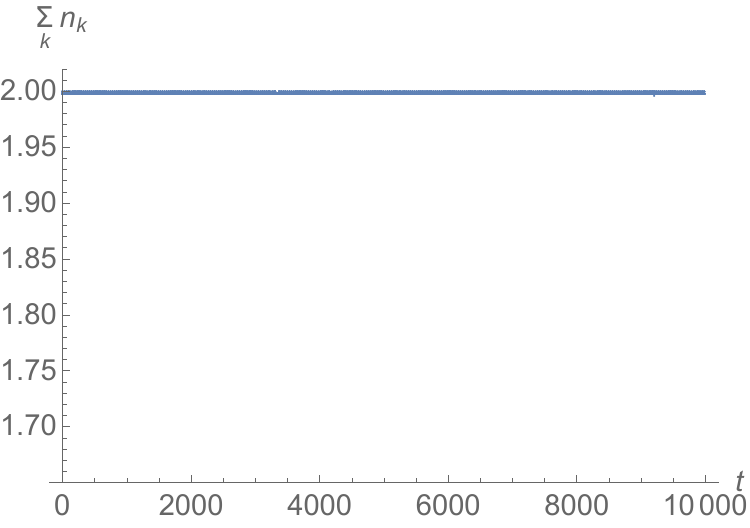}
		\caption{$C_{\text{m}} = 0.1$.} 
		\label{sfig:genericEvolution}
	\end{subfigure}
	\begin{subfigure}{0.4\textwidth}
		\includegraphics[width=\textwidth]{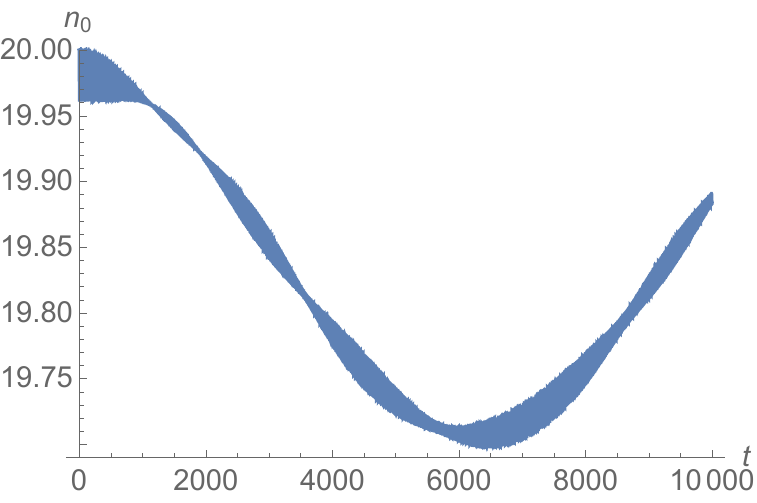}
		\hspace{0.18em} 
		\includegraphics[width=\textwidth]{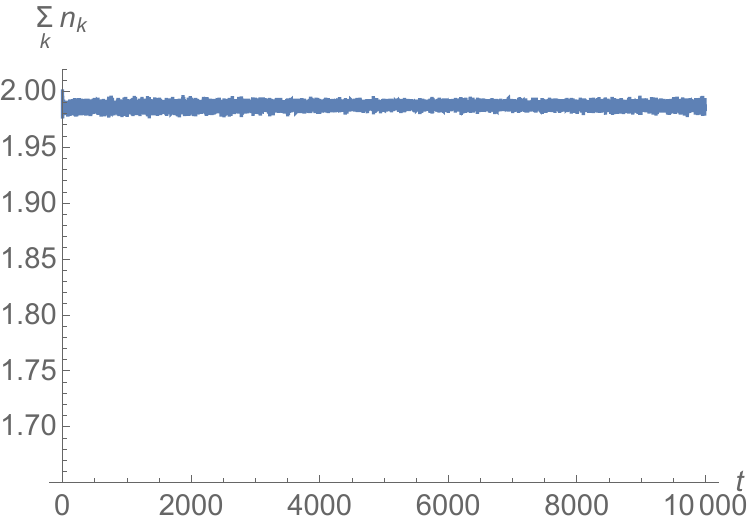}
		\caption{$C_{\text{m}} = 0.30055$. 
		} 
		\label{sfig:featureAEvolution}
	\end{subfigure}
	\hspace{0.05\textwidth}
	\begin{subfigure}{0.4\textwidth}
		\includegraphics[width=\textwidth]{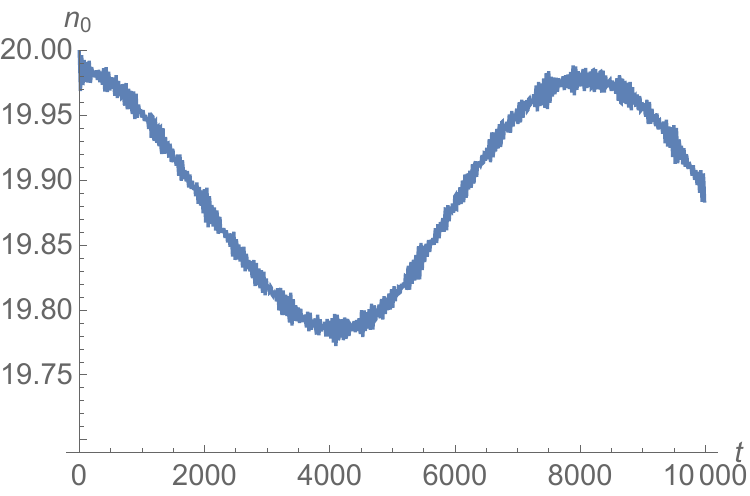}
		\hspace{0.18em} 
		\includegraphics[width=\textwidth]{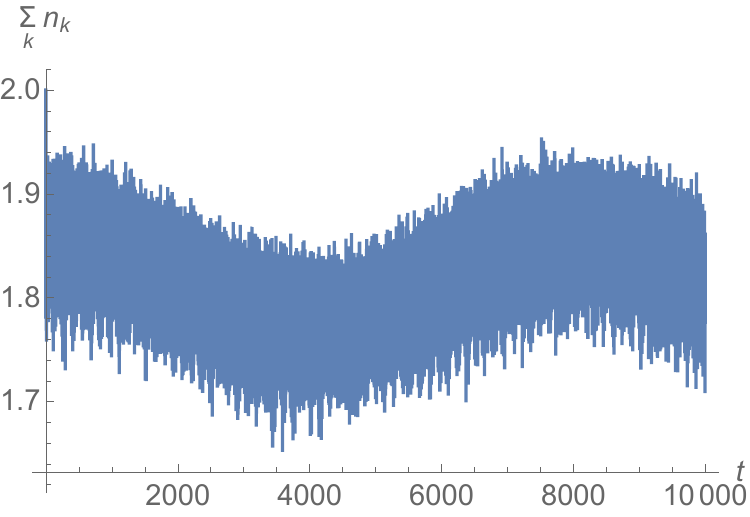}
		\caption{$C_{\text{m}} = 1.239$.} 
		\label{sfig:featureBEvolution}
	\end{subfigure}
	\caption{Time evolution of the initial state \eqref{initialStateSimple} for different values of $C_{\text{m}}$. Oscillations on a timescale of order $1$ cannot be resolved graphically any more since we show very long timescales. $n_0$ is the expectation value of the occupation of the mode $\hat{a}_0$ and $\sum_k n_k$ that of the total occupation in the first critical sector. Time is plotted in units of $\epsilon_0^{-1} \hbar$.}
	\label{fig:realTimeEvolution}
\end{figure*}

The time evolution of the initial state \eqref{initialStateSimple} for different values of $C_{\text{m}}$ is displayed in \fig \ref{fig:realTimeEvolution}, where we show the expectation value $n_0$ of the occupation number of the $\hat{n}_0$-mode as well as the expectation value of the total 
occupation of the first critical sector $\sum_{k=1}^K \hat{n}_k$. 
For $C_{\text{m}}=0$ (see \fig \ref{sfig:freeEvolution}), 
we can replace $\sum_{k =1}^{K}  \hat{n}_k \rightarrow N_m$ and $\sum_{k' =1}^{K'}  \hat{n}'_{k'} \rightarrow 0$ and the system has the analytic solution \eqref{expectationValueAnalytic}. We observe that the critical sector does not move and the amplitude of oscillations of $\hat{n}_0$ is strongly suppressed. 
This is the effect of memory burden \cite{1810.02336} discussed before in 
section \ref{ssec:memoryBurden}.

For many nonzero values of $C_{\text{m}}$, the system behaves similarly (see \fig \ref{sfig:genericEvolution}). Although the time evolution of the 
system becomes more involved, the amplitude of oscillations of $n_0$ is still small and the critical sectors remains effectively frozen. 

However, there are certain values of $C_{\text{m}}$ for which the system behaves qualitatively differently and the amplitude of oscillations of $n_0$ increases distinctly, albeit on a significantly longer timescale (see 
\figs \ref{sfig:featureAEvolution}, \ref{sfig:featureBEvolution}).
As expected, this behavior is accompanied by a change of the occupation numbers in the critical sector. This can either happen via an instantaneous jump (as in \fig \ref{sfig:featureAEvolution}) or via oscillations that 
are synchronous with $n_0$ (as in \fig \ref{sfig:featureBEvolution}). Although the second scenario is more intuitive than the first one, both are in line with our statement that $n_0$ can only change significantly if also rewriting in the critical sector takes place.
Moreover, we note that the occupation transfer and thus the rewriting of information is not complete.
We expect that complete rewriting into the second sector of memory modes can be achieved only after including further sectors, to which the $\hat{a}'_{k'}$-modes can transfer occupation number.

\begin{figure*}
	\begin{subfigure}{0.4\textwidth}
			\includegraphics[width=\textwidth]{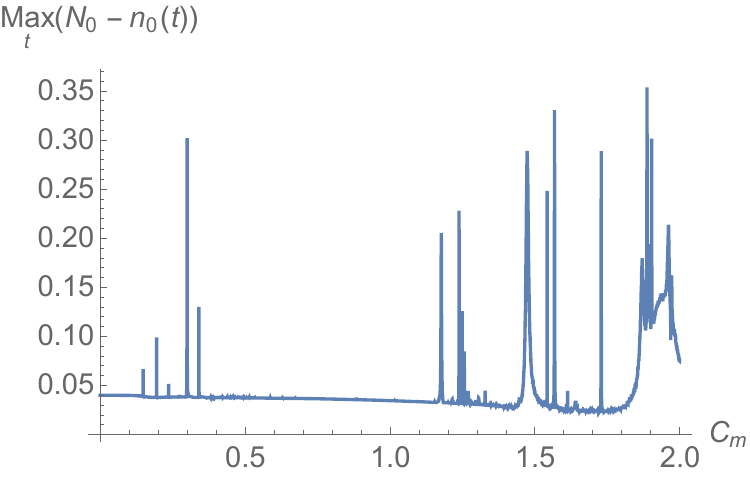}
		\caption{$p=1\,,~  C_0 = 0.01$}
		\label{fig:maxAmplitude}
	\end{subfigure}
	\hspace{0.05\textwidth}
		\begin{subfigure}{0.4\textwidth}
	\includegraphics[width=\textwidth]{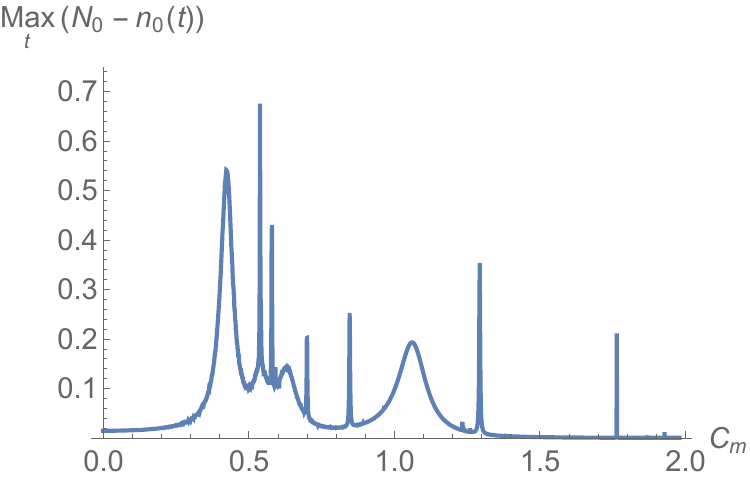} 
		\caption{$p=2\,,~  C_0 = 0.0003$}
		\label{fig:maxAmplitudep2}
	\end{subfigure}
	\caption{Maximal amplitude of the expectation value of $\hat{n}_0$ for different values of $C_{\text{m}}$ (with initial state \eqref{initialStateSimple}).}
\end{figure*}

We shall call the values of $C_{\text{m}}$ for which partial rewriting takes place \textit{rewriting values}.
For the present values of the remaining parameters \eqref{systemParameters}, such values are rare. In order to illustrate this point, we plot the maximal amplitude of oscillations as a function of $C_{\text{m}}$ in \fig 
\ref{fig:maxAmplitude}. 
We remark, however, that for other parameter choices, we have observed much more abundant rewriting values.

Finally, we also study how the system behaves when we choose $p=2$ (see 
\Eq \eqref{HamiltonianHigherOrder}), \ie we replace $1-\hat{n}_0/N_c \rightarrow (1-\hat{n}_0/N_c)^2$ and $1-\hat{n}_0/(N_c-\Delta N_c)\rightarrow 
(1-\hat{n}_0/(N_c-\Delta N_c))^2$. As is evident from \Eq \eqref{smallerBurden}, this reduces the memory burden by a factor of approximately $0.03$ (we used that $N_c-n_0$ can get as large as $0.3$ in the case $p=1$; see \fig \ref{fig:maxAmplitude}). In order to keep the amplitude of oscillation in the absence of rewriting (see \Eq \eqref{expectationValueAnalytic}) on the same order of magnitude as for $p=1$, we reduce $C_0$ by the same factor, \ie set $C_0 =0.0003$. Apart from this change, we use the values displayed in \Eq \eqref{systemParameters}. As in the case $p=1$, we time evolve the system for different values of $C_{\text{m}}$. The resulting maximal amplitude as a function of $C_{\text{m}}$ in displayed \fig \ref{fig:maxAmplitudep2}. Qualitatively, we observe the same behavior as for $p=1$.

\subsection{Dependence on System Size}
After having seen that rewriting does indeed take place, we now wish to answer the question how this changes as the size of the system increases. In order to answer it, we investigate how the rewriting values of $C_m$ change as we vary the parameters of the Hamiltonian \eqref{fullHamiltonianSimple}. 
Moreover, it is possible to determine a rate $\Gamma$ of rewriting as the 
ratio of the maximal amplitude of $n_0$ and the timescale on which this maximal value is attained. If we map our system on a decay process, as we have e.g., done in section \ref{ssec:blackHole}, we can identify this rate with the decay rate. We shall also study how the rate $\Gamma$ changes as we vary the parameters. In the following, we restrict ourselves to the 
case $p=1$. All numerical data that lead to the results shown subsequently are publicly available in the companion Zenodo record \cite{data}.

As presented in the appendix, the observed scalings for $C_m$ and $\Gamma$ are as follows:
\begin{itemize}
	\item The initial occupation number $N_c$ of $\hat{n}_0$ (which is also the critical occupation at which the first memory sector is gapless):
	\begin{equation}\label{NDependence}
		C_{\text{m}} \sim  N_c^{-1}\,, \qquad \Gamma \sim N_c^{-1} 
	\end{equation}
	\item The free gap $\epsilon_m$ of the memory modes:
	\begin{equation}
	C_{\text{m}} \sim \epsilon_m^1 \,, \qquad \Gamma \sim \epsilon_m^0 \ \text{(independent)} 
	\end{equation}
	\item The coupling $C_0$ of $\hat{a}_0$ and $\hat{b}_0$:
	\begin{equation}\label{C0Dependence}
	C_{\text{m}} \sim C_0^{0}\ \text{(independent)}  \,, \qquad \Gamma \sim C_0^{1.4}
	\end{equation}
	\item The difference $\Delta N_c$ between the critical occupations of $\hat{a}_0$ making either of the two memory sectors gapless:
	\begin{equation}
	C_{\text{m}} \sim (\Delta N_c/N_c)^{0.2}  \,, \qquad \Gamma   \sim (1-\Delta N_c/N_c)
	\end{equation}
\end{itemize}

\begin{figure*}
	\begin{subfigure}{0.45\textwidth}
		\includegraphics[width=\textwidth]{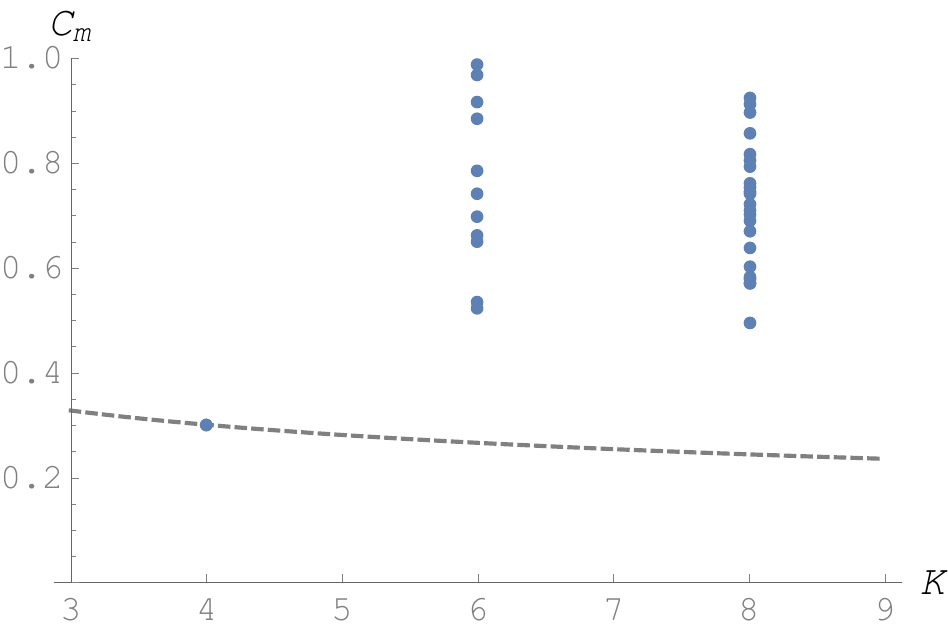}
		\caption{Rewriting values of $C_{\text{m}}$
		} 
		\label{sfig:QDependenceCm}
	\end{subfigure}
	\hspace{0.05\textwidth}
	\begin{subfigure}{0.45\textwidth}
		\includegraphics[width=\textwidth]{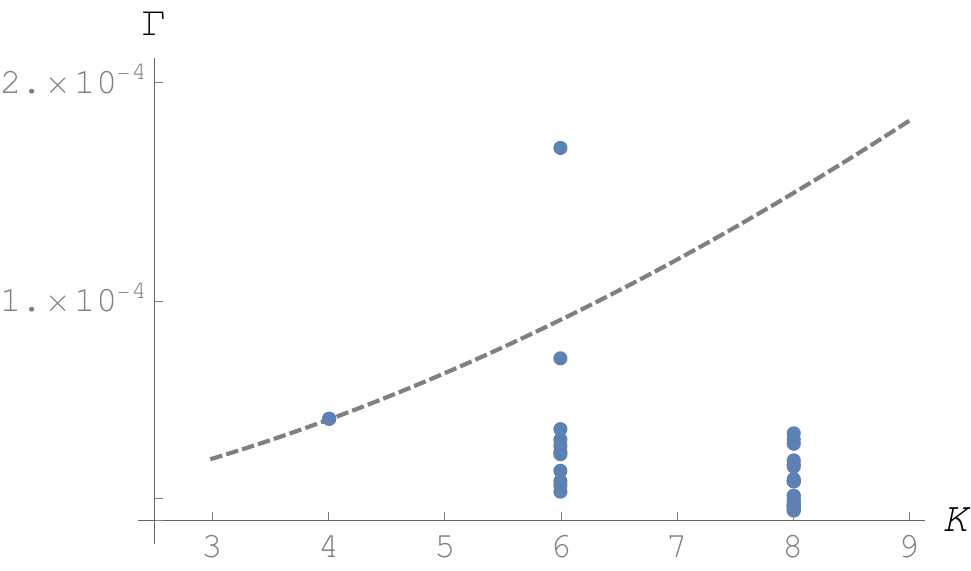}
		\caption{Rates $\Gamma$ at those rewriting values}
		\label{sfig:QDependenceRate}
	\end{subfigure}
	\caption{Available data (blue dots) for the rewriting values of $C_m$ and the rates $\Gamma$ as function of $K=K'$, where we take $N_m = K/2$.
 The dashed curves are the constraints \eqref{betaCBH} and \eqref{betaGammaBH}, that apply to a black hole. We see clear indications that for large black holes, rewriting is not fast enough to reproduce the semiclassical rate of evaporation.}
	\label{fig:QDependence}
\end{figure*}

With regard to the $K$ and $K'$, we can unfortunately only study three values due to numerical limitations, namely $K=K'=4,6,8$. The results are displayed in \fig \ref{fig:QDependence}, where we take $N_m = K/2$. Since we cannot make a precise statement about the dependence on $K$, we will parameterize it as 
	\begin{equation}
 C_{\text{m}} \sim K^{\beta_C}  \,, \qquad \Gamma   \sim  K^{\beta_\Gamma} \,.
 \label{fitQ}
 \end{equation}

 Still, we can try to constrain rewriting, \ie give a lower bound on $C_m$ and an upper bound on $\Gamma$. 
		To this end, we calculate the mean value of $C_m$ for the 11 data points at $K=6$. In order to obtain a maximally conservative bound, we moreover choose among the results for $K=8$ the 11 data points with the lowest values of $C_m$ and compute their mean. Performing a fit with the two resulting mean values, we get
		\begin{equation} \label{QDependenceEstimateCm}
		\beta_C \gtrsim -0.7 \,.
		\end{equation}
		We have not included the value at $K=4$ since doing so would increase 
$	\beta_C$.
		 For $\Gamma$, we compute the mean value for the 11 data points at $K=6$. At $K=8$, we choose the 11 data points with the highest rates and calculate their mean. Fitting the resulting two means together with the rate of the single rewriting value at $K=4$, we arrive at
		 \begin{equation} \label{QDependenceEstimateRate}
		\beta_\Gamma \lesssim -0.7\,.
		 \end{equation}
		 In this case, we have not excluded the value at $K=4$ because this would decrease $\beta_\Gamma$.
Even though we have tried to be conservative in our estimates, we must stress that we have far too little data to make any reliable statement. Thus, the true values of $\beta_C$ and $\beta_\Gamma$ might not respect the bounds \eqref{QDependenceEstimateCm} and \eqref{QDependenceEstimateRate}.

In summary, we observe that the rate $\Gamma$ decreases as the system size increases, \ie as $N_c$ or $K$ gets bigger (see \Eqs \eqref{NDependence} and \eqref{QDependenceEstimateRate}). Whereas $C_0$ can vary independently in a generic system, \Eq \eqref{boundC0} shows that it decreases with 
system size in the black hole case. Therefore, the observed scaling \eqref{C0Dependence} reinforces the tendency of lowering $\Gamma$. Thus, we see clear indications that rewriting becomes more difficult for larger systems.

 \subsection{Understanding Our Results} 

In this chapter, we shall provide some analytic understanding of our findings. 
For simplicity we take $K=K'$,  and assume each memory 
set to be diagonal  with the equal gaps. Thus, we take $\tilde{C}_{k,l} = 
\tilde{C}_{k',l'}= 0$ and  $\epsilon_k$ and $\epsilon_{k'}$ to be universal, i.e.,  for all $k,k'$, 
$\epsilon_{k} = \epsilon_{k'} = \sqrt{N_c}\epsilon_0$.  
Then, without any loss of generality, the matrix  
$C_{k,k'}$ can be set diagonal since we can always achieve this by an unitary transformation.
The Hamiltonian  (\ref{fullHamiltonian}) then becomes,  
\begin{align}
\hat{H} &= \epsilon_0 \hat{n}_0 + \epsilon_0 \hat{m}_0  +  C_0\left( \hat{a}_0^\dagger \hat{b}_0 + \hat{b}_0^\dagger \hat{a}_0\right) 
\nonumber\\
& +  {\mathcal E} \sum_{k}  \hat{n}_k
+  {\mathcal E}' \sum_{k}  \hat{n}_k'  
\nonumber\\
& + \sum_{k}  C_{k,k}\left( \hat{a}_k^\dagger \hat{a}'_{k} +  \text{h.c.}\right) \,, 
\nonumber\\
\label{SimpleHamiltonian}
\end{align}
where 
\begin{equation} 
{\mathcal E} \equiv \sqrt{N_c}\left(1-\frac{\hat{n}_0}{N_c}\right)\epsilon_0\,, 
\end{equation}
and 
\begin{equation} 
{\mathcal E}^{'} \equiv   \sqrt{N_c}\left(1-\frac{\hat{n}_0}{N_c-\Delta N_c}\right)\epsilon_{0} \,.
\end{equation}

Now,  in the initial state (\ref{initialState}), 
$n_0=N_c$ and we have 
\begin{equation} 
{\mathcal E}  = 0 \,,~  {\mathcal E}^{'} \simeq - \frac{\Delta N_c}{\sqrt{N_c}}\epsilon_{0}  < 0\, .
\end{equation}
Since the second gap is negative, there exist states 
with much lower energy than the initial one
(\ref{initialState}).  In particular, such is the state 
\begin{equation} \label{low}
\ket{\text{low}}=\ket{\underbrace{N_c}_{n_0},\underbrace{0}_{m_0},0, \ldots 0, \underbrace{n_1}_{n'_1}, \ldots, \underbrace{n_K}_{n'_{K}}} \,.
\end{equation}
This is the state obtained from (\ref{initialState}) by exchanging the occupation numbers of the two memory sectors, 
$n_k \leftarrow \rightarrow n_k' $, without touching 
$n_0$ and $m_0$.  
This state has a macroscopically large negative 
energy difference with respect to (\ref{initialState}). 
Expressing this energy difference 
in terms of the memory burden, which in this case implies    
$\epsilon_0 \sum_k n_k = - \mu \sqrt{N_c}$,  we have, 
\begin{equation} 
\bra{\text{low}} \hat{H} \ket{\text{low}} 
- \bra{\text{in}} \hat{H} \ket{\text{in}}
= {\mathcal E}' \sum_k n_k \sim 
\Delta N_c \mu \,. 
\end{equation} 
Since $\mu < 0$, 
the energy of  (\ref{low})
is much lower than the energy of the initial state (\ref{initialState}). 

The above makes it clear that we should be able 
to take intermediate deformations of the initial state
(\ref{initialState})
in which 
the exchange of the occupation number between the sets 
$n_k$ and $n_k'$ is balanced by the exchange
between $n_0$ and $m_0$ in such a way that the 
obtained state is nearly degenerate with the initial one. 

It therefore looks plausible that the system can use such a path 
for overcoming the memory burden. 
This was our original expectation as well as the proposal in 
\cite{1810.02336, 1812.08749}. 
Yet, this is not what we are finding. 

Instead we discover that the system cannot evolve efficiently. 
The reason for this is the following. 
The maintenance of the zero energy balance along the 
current trajectory requires a synchronous evolution of the two 
sets of degrees of freedom,  $a_0,b_0$ and $a_k,a_k'$, respectively. 
These sets have to produce opposite contributions  
into the energy that cancel each other.  
Basically, one set has to ``climb up" the energy stairs while the other 
is coming down.  

However, each evolution has a highly suppressed amplitude due to 
the huge gap differences among the proper modes. 
This breaks the process.  

In order to see this, let us consider the time evolution near the 
initial state  (\ref{initialState}). This time evolution can be described
as set of coupled $2\times 2$ problems, with the following Hamiltonians, 
\begin{equation} \label{mass2}
\hat{H}= \sum_k \bordermatrix{
	& \hat{a}_k	& \hat{a}_k^{'} 	\cr
	\hat{a}_k^{\dagger}	& 0	& C_{k,k} 	\cr
	\hat{a}_k^{'\dagger}	& C_{k,k} 	& {\mathcal E}'	\cr
} + 
\bordermatrix{
	& \hat{a}_0	& \hat{b}_0 	\cr
	\hat{a}_0^{\dagger}	& \epsilon_0 + \mu	& C_{0} 	\cr
	\hat{b}_0^{\dagger}	& C_{0} 	& \epsilon_0	\cr
}  \,.
\end{equation}
The systems are coupled because ${\mathcal E}'$ is 
a function of $n_0$, whereas $\mu$ is a function of
$\sum_k n_k$.  

The qualitative behavior can be understood  by solving the system iteratively, \ie in the zeroth order approximation, we evolve 
the system by treating  ${\mathcal E}'$ and $\mu$ as constants. 
We then evolve them in the first order by taking into account the variations 
of the occupation numbers obtained in the zeroth order.

It is immediately clear that the  amplitudes of both transitions 
are highly suppressed because of the huge eigenvalue-splittings in 
the corresponding $2\times 2$ matrices.
Namely, we have
\begin{equation} \label{changes} 
\delta n_0 \sim - N_c \frac{C_0^2}{\mu^2} \, ~~ {\rm and} ~~
\delta n_k \sim - \frac{C_{k,k}^2}{ {\mathcal E}^{'2}} \,. 
\end{equation} 
The resulting variations of ${\mathcal E}^{'}$ and $\mu$  are so small 
that they cannot affect the picture in the next iteration. 
For example, we can assume $K=N_c = \sum_k n_k$ and take into account
the bound $C_0 \lesssim \epsilon_0/\sqrt{N_c}$, which arises in the black 
hole case (see \Eq \eqref{boundC0}), as well as condition \eqref{epsilonEffCondition},\footnote
{We note that the stronger condition \Eq \eqref{boundCmBH} does not apply 
in the present case since the memory modes only couple pairwise.}
\begin{equation} \label{Cless}
\frac{C_{k,k}^2}{  
	{\mathcal E}^{'}}  \lesssim \frac{\epsilon_0}{\sqrt{N_c}}\,.
\end{equation}
Then we get
\begin{equation} 
\frac{\delta \mu}{\mu} \lesssim \frac{\epsilon_0}{\sqrt{N_c} {\mathcal E}^{'}}\, ~~
{\rm and}~~ \frac{\delta {\mathcal E}^{'}}{{\mathcal E}^{'}} 
\sim  \frac{1}{\Delta N_c N_c} \,.    
\end{equation}
Finally, we take into account that there is a lower bound on $|{\mathcal E}^{'}|$ (\Eq \eqref{boundEpsilonPrime} or \eqref{boundEpsilonPrimeStronger}). Even the mildest bound \eqref{boundEpsilonPrime} suffices to conclude that the iteration series  rapidly converge
towards the result that the system is essentially trapped in the initial state.    
Moreover, we note that \eqref{Cless} acquires an additional factor $1/\sqrt{N_c}$ if the effective gaps do not come with both positive and negative signs (see footnote \ref{footnote:random}).

\subsection{Application to Black Hole}
\label{ssec:numericalBH}

Now we adapt the parameter choice \eqref{parametersBH} that corresponds to the case of a black hole.
In this situation, only the couplings $C_0$ and $C_{\text{m}}$ as well as 
$\Delta N_c$ remain independent of $S$. However, $C_0$ is related to $S$ through the bound \eqref{boundC0}, which is specific to black holes. Likewise, \Eq \eqref{boundEpsilonPrime} leads to a bound on $\Delta N_c$, namely
\begin{equation}
	\Delta N_c \gg 1 \,.
\end{equation}
 Using the observed scalings stated in \Eqs \eqref{NDependence}-\eqref{C0Dependence}, we get for $C_{\text{m}}$
\begin{equation}\label{couplingBH}
C_{\text{m}}  \sim S^{-0.5 + \beta_C} (\Delta N_c/S)^{0.2} \gtrsim S^{-0.7 + \beta_C} \,,
\end{equation}
as well as\footnote
{Since $\Delta N_c/S \rightarrow 0$, the rate $\Gamma$ becomes independent of $\Delta N_c$.}
\begin{equation} \label{rateBH}
\Gamma \sim S^{-1.7 + \beta_\Gamma} \,.
\end{equation}
\Eq \eqref{couplingBH} shows that in order to satisfy the bound \eqref{boundCmBH} on the $S$-dependence of $C_m$, the scaling of $C_m$ with $K$ would be constrained as
\begin{equation} \label{betaCBH}
	\beta_C \lesssim -0.3 \,.
\end{equation}

Analogously, it follows from \Eq \eqref{rateBH} that the requirement of reproducing the semiclassical value of the rate, $\Gamma \sim 1$, leads to 

\begin{equation} \label{betaGammaBH}
	\beta_{\Gamma} \gtrsim 1.7 \,.
\end{equation}

Now we would like to investigate the compatibility of the numerical results for $K$-variation with the bounds in \Eqs \eqref{betaCBH} and \eqref{betaGammaBH}.
Firstly, we compare the actual results for $K=6,8$ with the expectation 
for $K=6,8$ based on the result for $K=4$ and a scaling saturating the bounds \Eqs \eqref{betaCBH} and \eqref{betaGammaBH}.
The resulting functions are plotted in \fig \ref{fig:QDependence}. We observe that although many more rewriting values exist at higher $K$, none of them satisfies both the constraints \eqref{betaCBH} and \eqref{betaGammaBH}.\footnote
{In fact, none of them fulfills either condition, except for one data point at $K=6$. It has a sufficiently high rate, but its coupling strength 
$C_m = 0.74$ is far too big to satisfy the bound \eqref{betaCBH}.}

As a different method of analysis, we can compare the bounds \eqref{betaCBH} and \eqref{betaGammaBH} with the estimates \eqref{QDependenceEstimateCm} and \eqref{QDependenceEstimateRate}. We observe that $\beta_C$ might be small enough, but $\beta_\Gamma$ is vastly different.
 These are clear indications that for large black holes rewriting of information from one memory sector to another one cannot be efficient enough to reproduce the semiclassical rate of particle creation, $\Gamma \sim 1$. 
As far as we can numerically access the system, we therefore conclude that even if rewriting takes place, the semiclassical description breaks down as soon as memory burden sets in.

We expect the semiclassical approximation to be valid as long as a newly created black hole has not lost a sizable fraction of its mass. In the absence of sufficiently fast rewriting, the model \eqref{fullHamiltonianSimple}, which we have studied here, does not fulfill this requirement because it would lead to an immediate deviation from the semiclassical rate of 
particle production.
Therefore, an effective description of black hole evolution must realize an appropriate delay of the onset of the memory burden effect. As reviewed in \ref{ssec:memoryBurden}, the model \eqref{HamiltonianHigherOrder} (with a parameter choice $p \gg 1$) achieves such a delay. It can do so at most until the master mode has lost on the order of half of its initial occupation, and, correspondingly, until the black hole has lost on the order of half of its mass.

We can give a quantitative estimate of how strong the slowdown is at that 
point, assuming that the rewriting rate after the onset of backreaction in the system \eqref{HamiltonianHigherOrder} (with a parameter value $p \gg 1$ and including the coupling to another set of memory modes) behaves analogously to the system investigated here.
To this end, we first note that $C_0\sim 1/S$ is required in order to reproduce the semiclassical rate of Hawking evaporation, $\Gamma \sim 1$, during the initial evolution before the onset of memory burden.
Consequently, \Eq \eqref{rateBH} gets modified:
\begin{equation} \label{rateBH2} 
\Gamma \sim S^{-2.4 + \beta_\Gamma} \,.
\end{equation}

As explained, we cannot determine $\beta_\Gamma$ due to numerical limitations. Still, we can try to give a bound on it. Since we clearly see no indications that the rates increase with $K$ (see \Eq \eqref{QDependenceEstimateRate}), we can conservatively estimate that $\beta_\Gamma<0$. Then we obtain 
\begin{equation}  \label{suppressedRate}
\Gamma \lesssim \frac{1}{S^2} \,.
\end{equation}
Thus, evaporation has to slow down drastically at the latest after the black hole lost on the order of half of its initial mass.

\subsection{Regime After Metamorphosis} 
In the following, we will discuss scenarios of BH evolution beyond half-decay that are consistent with the above finding.
 In the standard semiclassical treatment,  
 the evaporation process of a black hole is taken to be {\it self-similar}, \ie it is assumed to be well described simply by a time-dependent mass 
$M(t)$ which in each moment of time determines the 
Schwarzschild radius and the temperature as $r_g = 2G_NM(t)$ and $T = 
(8\pi G_NM(t))^{-1}$, respectively. 
 This leads to the picture of a thermal emission spectrum shifting with the growing temperature as the evaporation proceeds.
Thus, the assumption is that a {\it classical} black hole with each 
{\it quantum} 
emission evolves into a {\it classical}  black hole of a lower mass. 
This picture is widely accepted,  despite the fact that there exist no self-consistent
semi-classical calculation giving such a time evolution. 

It is quite contrary \cite{1509.04645}: This picture has a built-in measure of its validity since 
the above equation, together with $\dot{M} \sim r_g^{-2}$, leads to $\dot{T}/T^2 \sim 1/S$.   
This quantity sets the lower bound on the deviation from thermality.
It vanishes only in strict semi-classical limit 
$G_N \rightarrow 0, ~M \rightarrow \infty,~r_g=\text{finite}$. 
It is important to note that in this limit $S \rightarrow \infty$. Therefore, 
the standard Hawking result is exact.  However, for finite mass black holes 
and non-zero $G_N$, the deviations from thermal spectrum 
are set by $1/S$.   

Already this fact tells us that it is unjustified to use the 
self-similar approximation over timescales 
comparable with black hole half-decay, $\tau \sim Sr_g$. Indeed, 
without knowing the microscopic quantum theory, one can never
be sure that the semi-classical approximation is not invalidated 
due to a build-up of quantum back-reaction over the 
span of many emissions.    

 As microscopic theory tells us \cite{1112.3359, Dvali:2013eja,Dvali:2012rt,Dvali:2012wq}, this is exactly what is happening: At the latest by the 
time a black hole loses of order of half of its mass, 
 the back-reaction is so strong that the semiclassical treatment can no longer be used.\footnote
 {Self-similarity is only recovered in the semiclassical limit $N_c\rightarrow \infty$ \cite{1203.3372, 1504.04384}.}
   In particular, the remaining black hole state is fully entangled after 
losing on the order of half of its constituents.

   The present study reveals a new microscopic meaning of the 
quantum back reaction. 
  Namely,  being states of maximal memory capacity, 
the black holes are expected to share the universal property  
of memory burden. Due to this phenomenon, the black hole evaporation rate 
must change drastically after 
losing half of its mass. What happens beyond this point 
can only be a subject to a guess work. 
However, given the tendency that  the memory burden 
resist the quantum evaporation, the two possible outcomes are: 1) 
A partial stabilization by slowing down the evaporation; 2) Classical disintegration into some highly non-linear gravitational waves.  
The second option becomes possible because after the breakdown of the description in terms of a classical black hole, we cannot exclude any more that the system exhibits a classical instability. Obviously, there could be a combination of the two options, where a prolonged period of slow evaporation transits into a classical instability. In the following, we shall 
focus on the first option as being the most interesting for 
the dark matter studies. 

Thus, motivated from our analysis of the prototype model, 
 we shall adopt that the increased lifetime due to the slowdown is
\begin{equation}\label{lifetimeSlowdown}
\tilde{\tau} \gtrsim r_g S^{1+k} \, ,
\end{equation}
where $k$ indicates the power of additional entropy suppression of the decay rate as compared to the semiclassical rate, $\Gamma \sim r_g^{-1}$. 
Although the spectrum is no longer thermal, we shall assume that the 
 mean wavelength of quanta emitted during this stage is still on the order of the initial Schwarzschild radius $\sim r_g$, as long as the mass is still on the order of the initial mass. We must stress, however, that we cannot exclude that the black hole starts emitting much harder quanta after memory burden has set in. In particular, as the gap increases, the memory modes become easier-convertible into their free counterparts. 
 	This conversion is likely a part of the mechanism by which the 
 	information starts getting released after the black hole's half decay.

\section{Role of Number Non-Conservation }  
\label{sec:NNC}

In our previous analysis we have explained the essence of the 
memory burden phenomenon and studied its manifestations 
on prototype systems numerically. In these prototype systems 
we have limited ourselves to particle number conserving interactions. 
One may wonder how important the role of particle number 
conservation is. In this section we wish to explain that it is not. 

In order to understand this, let us first briefly recount the essence of the phenomenon which is the following.  A set of memory modes, which we have denoted by 
$\hat{a}_k$-s, becomes gapless when the occupation number of 
the master mode, $\hat{a}_0$, reaches a certain critical value 
$n_0 = N_c$. This allows to populate the memory modes -- and 
therefore store the memory patterns of the form (\ref{Mpattern}) --  
without any expense in energy.  
This memory pattern creates a backreaction on  
a master mode by generating a high
energy gap for it.  We call this gap a memory burden  and denote 
it by the symbol $\mu$.

The net effect of this backreaction is that it resists to any departure of 
the master mode away from criticality. This is because the information 
pattern stored in the memory modes (\ref{Mpattern})
becomes costly in energy
as soon as the the master mode moves away from the critical value. 
The change of energy, resulting from the variation $\Delta n_0$ of the master mode, is: 
\begin{equation} \label{Pcost} 
\Delta E_{\rm pattern} =  \mu \, \Delta n_0 \, .
\end{equation} 
As a result, a force is created that prevents the occupation number of 
the master mode from changing.

In order to allow for such a change, the system must get rid of the 
memory burden $\mu$ by somehow decreasing the occupation number of the memory modes.  The pattern stored in these modes then must be rewritten 
in some other set of modes which we have denoted by $\hat{a}'_{k'}$.  
Our simulations indicate that for the reasonable choice of the parameter values, in particular motivated by scalings in black hole analogs,  
the rewriting process is not nearly efficient for avoiding the memory burden, \ie the process of offloading the 
memory modes from the $\hat{a}_k$-sector into $\hat{a}'_{k'}$ 
cannot be synchronized with the decay of the 
$\hat{a}_0$-mode.  

So far, in our examples, both processes were modeled by interactions that 
conserve the
total number of particles, \ie the destruction of an $\hat{a}_k$-mode 
was accompanied by a creation of a particle from another sector. 
Likewise, the destruction of an excitation of the master mode $\hat{a}_0$ 
was 
accompanied by a creation of a $\hat{b}_0$-quantum. 

Our analysis showed that the memory burden effect was so strong that it 
killed the decay process  before 
the extra sectors ($\hat{a}'_{k'}, \hat{b}_0$) 
had any chance to be significantly populated. 
Correspondingly, the inverse transition processes played no role
in the time-evolution of the system.  

In these circumstances, it is clear that the role of 
particle number conservation is insignificant. 
Namely, the transitions generated by the 
particle number non-conserving interactions of similar strengths 
would be equally 
powerless against the memory burden effect.  Let us
estimate this explicitly. \\

\subsection{Number non-conserving decay of the master mode} 

We shall first consider the effect of number non-conserving decays 
of the master mode. 
We take the fixed memory burden $\mu$ and replace the 
number conserving mixing between 
$\hat{a}_0$- and $\hat{b}_0$-modes in (\ref{HamiltonianBandA}) by the following 
number-non-conserving one,   
\begin{eqnarray}  \label{HHH} 
&&\hat{H}  = 
(\epsilon_0 + \mu) \hat{a}^{\dagger}_0 \hat{a}_0  \, + 
\epsilon_0 \hat{b}^{\dagger}_0 \hat{b}_0  + 
C_0 (\hat{a}_0 \hat{b}_0  + \hat{a}_0^{\dagger} \hat{b}_0^{\dagger})\, , 
\end{eqnarray}
where we have taken the parameter $C_0$ to be 
real and of the same strength as in the number-conserving version
(\ref{HamiltonianBandA}).  This strength, 
$C_0 \sim \epsilon_0/N_c$, is dictated by the requirement 
that for $\mu=0$ the half-decay time of the $\hat{a}_0$-mode is 
$t \sim N_c/\epsilon_0$  as it is clear from (\ref{expectationValueAnalytic}). 
This choice imitates, at the level of our toy model,  the scaling of the black hole  half-decay time  in units of the energy of the Hawking quanta.

The Hamiltonian (\ref{HHH}) is diagonalized by the following Bogoliubov transformation:
\begin{equation} 
\hat{a}_0 = u \hat{\alpha} - v \hat{\beta}^{\dagger} \,, \qquad 
\hat{b}_0 =  u \hat{\beta} - v \hat{\alpha}^{\dagger}  \,,   
\end{equation} 
where $\hat{\alpha}$ and $\hat{\beta}$ are the eigenmodes and 
\begin{align} \label{vu1}
v^2 &= \frac{1}{2} \left (\frac{1}{\sqrt{1 - \frac{4C_0^2}{(2\epsilon_0 
+ \mu)^2}}}- 1\right), \\
 u^2 &= \frac{1}{2} \left (\frac{1}{\sqrt{1 - \frac{4C_0^2}{(2\epsilon_0 + \mu)^2}}} +  1\right)\,.
\end{align} 
For the full memory burden, we have $\mu \sim -\epsilon_0 \sqrt{N_c}$. 
Taking into account that $C_0 \sim \epsilon_0/N_c$, \Eq
(\ref{vu1}) gives,  
\begin{equation} 
v^2 \simeq  \frac{C_0^2}{\mu^2} \sim  
\frac{1}{N_c^3}\,, ~~~
u^2 = 1 + {\mathcal O}(1/N_c^3)\,,  \label{depletionMasterMode}
\end{equation} 
Thus, the depletion coefficient $v^2$ is minuscule due to an extra 
$1/N_c$-suppression by the memory burden factor $\frac{\epsilon_0}{\mu}$.
 We note that the amplitude in
	the case of a number conserving mixing is suppressed by the same factor $C_0^2/\mu^2$ (see \Eq (\ref{expectationValueAnalytic})).

 The similarity between the two cases in the present situation, 
in which the occupation number of the $\hat{b}_0$-mode never becomes high 
whereas the occupation number of $\hat{a}_0$ is macroscopic, can also be understood in the Bogoliubov
approximation, in which we replace number operators of the 
$\hat{a}_0$-mode by $c$-numbers, $\hat{a}_0 = \sqrt{N_c}\,,~  
\hat{a}_0^{\dagger} = \sqrt{N_c}$. After disregarding terms 
smaller than $1/\sqrt{N_c}$,\footnote{
This is justified (self-consistently) as long as the departure of $\hat{a}_0$ from $\sqrt{N_c}$ is small.
}
 the number non-conserving Hamiltonian
(\ref{HHH}) becomes:
\begin{eqnarray}  \label{HNONB} 
&&\hat{H}  = 
\epsilon_0 \hat{b}^{\dagger}_0 \hat{b}_0  + 
C_0 \sqrt{N_c}(\hat{b}_0^{\dagger} + \hat{b}_0)\,.  
\end{eqnarray}
This is easily brought to a diagonal form by a canonical transformation 
\begin{equation}  
\hat{b}_0 =  \hat{\beta} - C_0 \sqrt{N_c}/\epsilon_0 \,,   
\end{equation} 
which shows that the occupation number of the $\hat{b}_0$-mode 
in the $\beta$-vacuum is, 
\begin{equation}
\langle \hat{b}_0^{\dagger} \hat{b}_0 \rangle =  
\frac{C_0^2 N_c}{\epsilon_0^2} \,.  
\end{equation}  
Thus, depletion is still suppressed by $C_0$. The above argument is another way of understanding why -- as long as  the occupation number of 
$\hat{b}_0$ is small and therefore the inverses processes are not effective --  
the number conserving versus number non-conserving 
nature of the mixing is not important.
Thus, it is clear that number non-conservation does not allow to circumnavigate the memory burden effect.

Further notice that the relative effect of possible 
decays of the $\hat{a}_0$-mode into several $\hat{b}_0$-quanta is 
negligible. In the interaction vertex each extra $\hat{b}_0$ brings an additional factor of
$1/\sqrt{N_c}$. So, the coefficient of the term 
$\hat{a}_0\hat{b}_0^{l}$ scales as $\epsilon_0/N_c^{l/2}$.   

\subsection{Number non-conserving decay of memory modes} 

Let us now ask whether 
a number non-conserving decay  of the memory modes 
could ease the memory burden effect more rapidly
as compared to a number conserving one.  
In order to answer this question, we shall  allow  for a depletion of the 
memory modes 
due to a number non-conserving mixing with another sector. 

  Before doing so we wish to clarify the following point. 
	Notice that for odd values of $p$ the effective gap 
	of the second layer of the memory modes becomes negative. 
	For a number conserving  Hamiltonian, this makes no difference 
	in the memory burden effect. 
	Therefore we can use the 
	simplest version of $p=1$ (see \eqref{effectiveGapHigherOrder}) in our 
discussions. 
	
	However, once we allow for number non-conservation among 
	memory modes, the negative gap can lead to ``tachyonic" type 
	instabilities that can populate the memory modes by creating them   
	out of the vacuum. This increase of the occupation number 
	has nothing to do with easing a memory burden on the first sector but 
	rather is a manifestation of instability.  Still, such instability 
	blurs the  question that we are after. 
	So we shall assume that the gap is always positive. 
	This assumption is equivalent 
	to the statement that there are no intrinsic tachyonic instabilities 
	in the initial state of the system. 
	This imposes a constraint on the structure of the Hamiltonian, which we will take into account below. 
	
	Let us consider the following Hamiltonian that mixes 
	the memory modes,    
	\begin{align}
	\hat{H} &= 
	{\mathcal E}_k \hat{n}_k
	+ {\mathcal E}_{k'}\hat{n}'_{k'} +
	\nonumber\\
	& +  C_{k,k'}\left( \hat{a}_k \hat{a}'_{k'} +  \text{h.c.}\right) \,.
	\label{HAA}
	\end{align}
	For simplicity of illustration,  we paired-up the modes from the two sets, 
	effectively reducing the evolution to a $2\times 2$ problem.  
	Making a more complicated 
	mixing matrix does not change the qualitative picture.
	
	The quantities ${\mathcal E}_k$ and ${\mathcal E}_{k'}$ are 
	effective gaps that are functions of $n_0$.  The only requirement that we need to specify is that they {\it alternatively} reach zeros
	for two critical values $n_0= N_c$ and $n_0=N_c - \Delta N_c$ respectively,  
	and are semi-positive definite everywhere in between.
	Moreover, at the critical points the level splitting is macroscopic:   
	\begin{align} \label{Cpoints}
	\nonumber
	{\mathcal E}_k = 0\,, {\mathcal E}_{k'} \gg \epsilon_0,     & \qquad\text{for} ~ n_0=N_c, 
	\\	
	{\mathcal E}_{k'} = 0\,, {\mathcal E}_{k} \gg \epsilon_0,       &\qquad \text{for} ~
	n_0=N_c-\Delta N_c.
	\end{align}
	For example, we can choose
	\begin{equation} 
	{\mathcal E}_k \equiv \left(1-\frac{\hat{n}_0}{N_c}\right)\epsilon_k\,, 
	\end{equation}
	as before, and for ${\mathcal E}_{k'}$ assume one of many possible shapes, e.g.,  
	\begin{equation} 
	{\mathcal E}_{k'} \equiv  \left(1-\frac{\hat{n}_0}{N_c-\Delta N_c}\right)^2\epsilon_{k'} \, ,
	\end{equation}
	or
	\begin{equation} 
	{\mathcal E}_{k'} \equiv \left(\frac{\hat{n}_0}{N_c-\Delta N_c}-1\right)\epsilon_{k'}\,,~ 
	\end{equation}
	and so on.\footnote{Even a gap function such as, e.g., ${\mathcal E}_{k'} \equiv  \left(1-\frac{N_c-\Delta N_c}{\hat{n}_0}\right)\epsilon_{k'}$ would be admissible since here we are not restricted by renormalizability and only interested in the regime of $n_0 \gg 1$.}
	 Finally, as in the number conserving case, to which we wish to compare the evolution of the present system, the parameter $C_{k,k'}$ cannot be too big since otherwise the gaplessness
		of $\hat{a}_k$-modes at the critical point $n_0=N_c$ would be destroyed. This leads to the condition \eqref{Cless}.

	Now, let us recall that the reason why the depletion of the gapless memory modes
	in the particle number-conserving case (\ref{fullHamiltonian}) was 
	not efficient,  is: 
	\begin{itemize}
		\item  Relatively large level-splitting: 
		$\Delta {\mathcal E} =  {\mathcal E}_{k'} - {\mathcal E}_{k}$
		\item Suppressed mixing coefficient $C_{k,k'}$. 
	\end{itemize}
	As long as these conditions are maintained,  
	allowing the particle number non-conservation in the mixing term does not improve the situation.    
	Indeed, we diagonalize the Hamiltonian via the Bogoliubov transformation,
	\begin{equation} 
	\hat{a}_k = u \hat{\alpha}_k - v \hat{\beta}_{k'}^{\dagger} \,,\qquad 
	\hat{a}_{k'}' =  u \hat{\beta}_{k'} - v \hat{\alpha}_k^{\dagger}  \,,   
	\end{equation} 
	where
	\begin{equation} 
	u^2 = 1+ v^2\,, ~ v^2 = \frac{1}{2} \left (\frac{1}{\sqrt{1 - \frac{4C_{k,k'}^2}{({\mathcal E}_k + \,  
				{\mathcal E}_{k'})^2}}}- 1\right) \,. 
	\end{equation} 
	Next, taking into account \Eqs \eqref{Cless} and (\ref{Cpoints}),  we have near $n_0=N_c$
	\begin{equation} 
	v^2 \sim  \frac{C_{k,k'}^2}{(  
		{\mathcal E}_{k'})^2} \ll \frac{\epsilon_0}{\sqrt{N} 	{\mathcal E}_{k'}}
	\,,\qquad\,
	u^2 \simeq 1\,, 
	\end{equation} 
	which gives a highly suppressed rate of offloading the memory pattern.
This rate is negligible and of no help for liberating 
the master mode $\hat{a}_0$ from the memory burden effect on 
any reasonable timescale.  

Finally, as in the previous example, the higher order operators, 
e.g.,   $\hat{a}_k\hat{a}_{k_1'}'\cdots \hat{a}_{k_l'}'$, 
(whether number conserving or not) cannot improve the decay rate due to 
the extra suppression by powers of $1/N_c$.  

\subsection{Summing up} 

We are now ready to summarize the discussion about the 
role of number (non-)conservation in easing the memory burden.  
The memory burden phenomenon is a {\it relative}  effect 
that amounts to a delay of the time evolution of the system
due to a backreaction of the memory pattern.  In order to measure its 
effect,  we must do so by comparing the time evolution of the
same system with and without the initial memory pattern.

Motivated by the black hole analogy,  we have normalized 
(in units of 
an elementary gap $\epsilon_0$)
the half-decay time of the system with $\mu=0$ (an empty memory pattern)  to $N_c$.  This timescale fixes the strengths of the leading operators 
that can reduce the number of the master mode. 

We then study the system, with maximal memory burden 
$\mu$ (fully loaded memory pattern), 
assuming the strict conservation of the occupation number of the 
memory  modes $\hat{a}_k$. 
In this case, we see that the memory burden stops
the leakage of the master mode at the latest after half-decay. 

The next question we have analyzed was what happens if the memory pattern 
-
initially loaded in memory modes $\hat{a}_k$ -- can be offloaded to another 
sector $\hat{a}'_{k'}$.  We call this process rewriting. 
Can the rewriting be so timely as to free the master mode
from the memory burden?  
The answer to this question turned out to be negative. 

The point is that the gaplessness of the memory modes 
at the critical point restricts the strengths of operators that 
change their occupation. 
This restriction is independent of the conservation of the total particle 
number.  
As a result, the particle number non-conserving interactions are not any more efficient than the number conserving ones.

Another physical way of understanding this is the following. 
In the initial state, the sectors $\hat{a}_0$ and $\hat{a}_k$ are fully loaded, whereas 
all the other sectors ($\hat{b}_0, \hat{a}_{k'}'$) are empty.  
Now the task is to offload the  $\hat{a}_k$-modes by using an interaction 
that can reduce their number, irrespective of the total particle number conservation.  
The only difference  is that  the particle number-conserving operators 
would create exactly equal number of quanta in the empty sectors, whereas 
the non-conserving ones would not. 
However, what we observe is that with the allowed strength of the operators, 
the empty sectors are never populated 
efficiently on the timescale of interest.  
So, on average, no inverse process plays any role. 
In this situation the total number conservation is irrelevant.   
For an observer in the $\hat{a}_k$-sector, there is simply no 
difference between how the memory modes decay, since the inverse processes 
are never seen.

\section{Small Primordial Black Holes as Dark Matter}
\label{sec:primoridalBlackHoles}

The possible stabilization of black holes by the burden of memory 
could have interesting consequences for the proposal that primordial black holes (PBHs) constitute dark matter (DM) \cite{PBH1,PBH2,PBH3,PBH4}.  Of course, the full investigation
of this parameter space requires more 
precise information about the behavior of black holes past their 
naive half life.  
 Below, we first give a short qualitative discussion of how some of the bounds on primordial black holes change in this case. Subsequently, we provide a few quantitative considerations for one exemplary black hole mass.

\subsection{Effects on Bounds}

There exist many different kinds of constraints on the possible abundance 
of PBHs (see \cite{1607.06077, 2002.12778} for a review). However, the strength and/or the range of many of those constraints are based on the semiclassical approximation for BH evaporation, \ie Hawking evaporation is assumed throughout the decay. Therefore, a slowdown due to the backreaction in form of memory burden, which sets in after the half-decay, affects the landscape of constraints quite dramatically.

If the validity of the semiclassical approximation is assumed throughout the whole decay process, all PBH with masses $M \lesssim M_* \equiv 5 \cdot 10^{14}g$ would have completely evaporated by the present epoch\cite{2002.12778}.  In contrast, such small PBHs can survive until today if evaporation slows down after half-decay. Thus, many of the constraints on the 
initial abundance of PBHs with masses $M \lesssim M_*$ are altered. In particular, a new window for PBHs as DM is opened up for some values of the 
mass below $M_*$.

For example, we can consider constraints from the galactic gamma-ray background,
following \cite{1604.05349}. 
Since the spectrum of photons observed due to PBHs clustering in the halo 
of our galaxy is dominated by their instantaneous emission, the range of the related constraints in the semiclassical picture applies to black holes with mass $M\gtrsim M_*$, with the strongest constraints coming from $M$ close to $M_*$ (since they would be in their final, high-energetic stage of evaporation today).
On the one hand, a slowdown significantly alleviates the constraint around $M_*$ since such black holes would now be in their second, slow phase of evaporation. On the other hand, because black holes with masses below $M_*$ could survive until today, the galactic gamma-ray background would lead to new constraints on their abundance. At the same time, the fact that these black holes emit energetic quanta opens up a possibility to search for them via very high-energetic cosmic rays. Below we discuss this point in more detail. 

As a different example, we consider constraints from big bang nucleosynthesis (BBN), as were studied 
in \cite{0912.5297}. In the semiclassical picture, PBHs of mass smaller than about $M_N \equiv 10^{10}\,\text{g}$ would have evaporated until then. Therefore, such black holes are typically considered to be unconstrained by BBN. In contrast, a slowdown would cause some PBHs with $M \lesssim M_N$ to still exist at that epoch. Therefore, BBN in principle leads to 
new constraints on such PBHs.
However, the constraints are expected to be mild, since PBHs would already be in their second, slow phase of evaporation.  On the other hand, the strong constraints on  $M \sim M_N$ associated with the final stage of evaporation in the Hawking-picture is alleviated. Finally, the bound due to 
BBN on PBHs of masses $M \gg M_N$ is the same in the semiclassical and our picture because those black holes are in the early stages of evaporation during BBN. 

\subsection{Specific Example}

In the following, we consider an exemplary scenario, in which small PBHs of mass below $M_*$ appear to be able to constitute all of dark matter. 
It should be clear that we make no attempt to cover the whole spectrum of 
constraints or the whole range of masses, and content ourselves with rough estimates.
We consider a monochromatic PBH mass spectrum with $M \sim 10^8 g$. Moreover, we need to specify how strong the slowdown is after half decay. Based on our numerical finding \eqref{suppressedRate}, we assume that the rate $\tilde{\Gamma}$ is suppressed by two powers of the entropy: $\tilde{\Gamma} \sim r_g^{-1}/S^2$. Correspondingly, we have $k=2$ in \Eq \eqref{lifetimeSlowdown}, \ie the lifetime $\tilde{\tau}$ is prolonged as $\tilde{\tau} \gtrsim S^2 \tau$, where $\tau$ is the 
 standard estimate based on extrapolation of 
Hawking's result.
This leads to $\tilde{\tau}\gtrsim 10^{49}\,\text{s}$ (see \cite{2002.12778} for $\tau$), which is longer than the age of the Universe by many orders of magnitude. 

There are two kinds of constraints on the PBHs that we consider. Bounds of the first type are independent of the fact that the PBHs evaporate, \ie 
they are identical to the ones for massive compact halo objects (MACHOs) of the same mass. We are not aware of relevant constraints for masses as low as $M \sim 10^8 g$ (see \eg 
\cite{2002.12778, astro-ph/9803082}).\footnote{Constraints would be similar to the ones on N-MACHOs \cite{1911.13281}.} The second kind of bounds is due to the fact that, although with a suppressed rate, the PBHs still evaporate. 

As explained above, the energy of emitted particles is expected to be around the initial black hole temperature, $T_{\text{BH}} = M_p^2/(8 \pi M) \sim 10^5\,\text{GeV}$. 
Assuming that the galactic halo is dominated by the PBHs, the diffuse galactic photon flux due to the PBHs can be roughly estimated as 
\begin{equation}\label{flux}
\Phi \sim n_{\text{BH}} \, R \, \tilde{\Gamma} \sim 10^{-34}/(\text{cm}^2 
\text{s}) \,,
\end{equation}
where $R\sim 2\cdot 10^{24}\,\text{cm}$ is the typical radius of the Milky Way halo and $n_{\text{BH}}$ is the galactic number density of PBHs. We 
can estimate the latter in terms of the mass of our galaxy $M_{\text{MW}} 
\sim 2\cdot 10^{42}\,\text{kg}$ as $n_{\text{BH}} \sim M_{\text{MW}}/(M R^3)$.
This corresponds to one particle hitting the 
surface of the earth approximately every $10^{8}$ years. Clearly, it is impossible to observationally exclude such a low flux.\footnote
{We are not aware of an observational lower bound on the diffuse galactic 
gamma-ray flux at photon energies $E_{\gamma} \sim 10^5\,\text{GeV}$. For 
$E_{\gamma} \sim 10^3\,\text{GeV}$, the observed flux is of order $10^{-10}/(\text{cm}^2 \text{s})$ \cite{1410.3696}.}
Moreover, one can wonder if the secondary flux, which predominantly comes 
from the decay of pions, can change the above conclusion. The answer is negative since the corresponding rate $\tilde{\Gamma}_S$ is only slightly higher than the one for primary emission, $\tilde{\Gamma}_S \sim 10\tilde{\Gamma}$ (see \cite{1604.05349}).

Moreover, we can turn to constraints from the extragalactic gamma-ray background. 
Assuming that cold DM is dominated by PBHs of mass $M$, one can roughly estimate for the flux due to secondary photons\footnote{The primary photons would effectively be screened by a cosmic gamma-ray horizon (see \eg \cite{gammaRayHorizon})} (see \cite{0912.5297}):
\begin{equation}\label{EGflux}
\Phi \sim  \, \frac{\rho_{\text{DM}}}{M}\tilde{\Gamma} \, t_0 \sim 10^{-31} /(\text{cm}^2 \text{s}) \, ,
\end{equation}
where $\rho_{\text{DM}} \sim 2\cdot 10^{-30}\,\text{g}/\text{cm}^3$ is the present energy density of dark matter in the Universe and $t_0 \sim  4\cdot 10^{17}\,\text{s}$ is the age of the Universe. 
Again, this flux is unobservably small.

Finally, the contribution from the considered PBHs to cosmic rays other than photons can be expected not to  exceed significantly the photonic flux, in which case no bound would result from direct detection of other particles, either.

In conclusion, from the exemplary constraints considered above, the numerical example of PBHs of mass $M \sim 10^8 g$ 
 passes an immediate test to be able to account for all DM. 
As stated above, a more complete analysis remains to be done.
 
 We finish the section by making a general remark. The stabilized black holes can be detected via their emission but also via a direct encounter with earth, through gravitational or seismic disturbance.  The latter possibility for standard PBH has been discussed in \cite{1203.3806}. In the present context, the encounter becomes much more frequent and for certain masses the detection through a direct encounter could in principle become 
more probable than by emission spectrum.\footnote{We thank Florian K\"uhnel for comments on this issue.}

\section{Summary and Conclusions}
\label{sec:outlook}	

\subsection{Stabilization by Memory Burden}
A state around which gapless modes exist possesses an enhanced capacity of memory storage since information patterns can be recorded in the excitations of the gapless degrees of freedom at a very low energy cost. However, the stored information backreacts on the evolution of the system and ties it to its initial state. This is the effect of memory burden \cite{1810.02336}.

In this paper, we have investigated if memory burden can be avoided once another set of degrees of freedom exists, which becomes gapless for a different state of the system and to which the stored information can be transferred. We refer to this process as rewriting. In a prototype model, we 
have found a positive answer. For certain values of the parameters, a non-trivial evolution in the form of rewriting is indeed possible. It turns out that the timescale of this process is very long and we have studied how it depends on the various parameters of the system.

We can choose the parameters of our prototype model in such a way that it 
reproduces the information-theoretic properties of a black hole, in particular its entropy. In this case, we have concluded that as far as we can numerically access the system, rewriting happens significantly too slowly 
to match the semiclassical rate of particle production. 
This strongly indicates that evaporation has to slow down drastically at the latest after the black hole has lost on the order of half its initial 
mass.

This could open up a new parameter space for primordial black holes as dark matter candidates. For sufficiently low masses, those black holes would evaporate on a timescale shorter than the age of the Universe if Hawking's semiclassical calculation were valid throughout their lifetime. It is 
often assumed that this is the case 
so that the corresponding mass ranges are considered as excluded.

By contrast, a significant slowdown of the rate of energy loss, as is \eg 
displayed in \Eq \eqref{suppressedRate}, allows the lifetimes of such PBHs to be much longer so that they can still exist today. In this case, small PBHs become viable dark matter candidates. We have qualitatively discussed how some of the constraints change and studied a concrete example. A 
full investigation of parameter space remains to be done.

Our findings also have interesting implications for de Sitter space. In \cite{1812.08749}, we have already discussed the role of memory burden for 
this system and how it leads to primordial quantum memories that are sensitive to the whole inflationary history and not only the last $60$ e-foldings. Since the information-theoretic properties of de Sitter are fully analogous to those of black holes, our results imply that avoiding memory burden by rewriting the stored information cannot be efficient for de Sitter, either. This further supports the conclusions of \cite{1812.08749}.

\subsection{Black Hole Metamorphosis After Half Decay}
Finally, our analysis adds paint to the quantum picture of black holes. It has been standard to assume that black hole evaporation 
is {\it self-similar} all the way until the black hole reaches the size of the cutoff scale. 
For example, after a solar mass black hole loses, say, $90$ percent of its initial mass, the resulting black hole is commonly believed to be indistinguishable from a 
young black hole with $0.1$ solar mass. 
  In other words, the standard assumption is that a black hole at any stage of its existence has no memory about its prior 
history. This assumption is based on naive extrapolation of Hawking's 
{\it exact}  semi-classical computation towards arbitrary late stages  
of black hole evaporation.  However, 
this extrapolation unjustly neglects the quantum 
back-reaction that alters the state of a black hole. 
The lower bound on the strength of the back-reaction effect can be derived using solely the self-consistency of Hawking formula
and is $\sim 1/S$ per each emission \cite{1509.04645}. 
This fact already gives a strong warning sign that we cannot extrapolate the semi-classical result over timescales of order $S$ emissions.  

 However, only in a microscopic theory such as the quantum 
 $N$-portrait \cite{1112.3359} it is in principle possible to account for 
back-reaction properly and to understand its physical meaning. 
 This theory predicts \cite{Dvali:2013eja,Dvali:2012rt,Dvali:2012wq,Dvali:2013vxa,Dvali:2017eba} that semiclassical treatment cannot be extrapolated beyond the point when the black hole loses half of its mass. 
  The physical meaning of this effect  is very transparent. 
 Indeed, at this point the black hole loses half of its graviton constituents that leave the bound state 
in form of the Hawking quanta.  The quantum state of the remaining 
$\sim N_c/2$ gravitons is fully entangled and is no longer representable as approximately-classical coherent state. 
 
 In the present paper, we have identified another very explicit  source of backreaction in form of the memory burden effect described earlier 
 in \cite{1810.02336, 1812.08749}. 
  This effect shows that black hole evaporation cannot be self-similar
  and that old black holes are drastically different from the younger  
  siblings of equal mass.  That is, the memory burden results in 
  a quantum hair that exerts a  strong influence at later stages. This provides a concrete mechanism for the release of information after Page's time \cite{hep-th/9306083}.
  Note, such a  quantum hair is not constrained by the 
  classical no-hair theorems \cite{
   nohairRuffiniWheeler, 
   Hartle:1971qq,
   Bekenstein:1972ny, 
   Bekenstein:1971hc,
   PhysRev.D5.2403, 
   Teitelboim:1972pk,
   Teitelboim:1972ps}.  
  
  While it is evident that the memory burden resists to quantum decay of a black hole,  the further evolution requires a more detailed
 study.  In particular, it is not excluded that some sort of a classical instability 
 can set in after half-decay.  Our prototype models are not powerful enough for 
 either capturing such instabilities or excluding them.
  However, they clearly indicate the tendency 
 of dramatic slow-down of the quantum decay. Our speculations 
 about the small black holes as dark matter candidates are based
 on this evidence.  Nevertheless, the possibility of developing a 
 classical instability after the suppression of the 
 quantum emission  remains feasible and must be investigated 
seriously. An interesting avenue in this direction would be 
an experimental simulation of the effect in laboratory, since the states of enhanced memory capacity
can be achieved in simple setups with cold bosons (see \cite{1805.10292} and references therein). 

 Another promising approach could be to use the recently established 
 connection between the saturation of the information storage capacity 
 and unitarity of scattering amplitudes in generic quantum field theories 
\cite{1906.03530, 1907.07332, 2003.05546}.  All studied saturated objects 
exhibit close similarities with 
 black holes and are also subjected to the effect of memory burden. 
 So the question about the classical instability versus stabilization 
 can be addressed by studying the latest stages of their time-evolution.

 In conclusion, the memory burden backreaction 
opposes the quantum decay and the resistance 
becomes maximal by the time the black hole loses half of its mass. 
This behavior and whatever happens after is very different from the standard semi-classical picture and changes our understanding 
of black holes.

\begin{acknowledgments}
It is a pleasure to thank Oleg Kaikov and Florian K\"uhnel for comments. We thank the anonymous referee for useful questions and suggestions.  This work was supported in part by the Humboldt Foundation under Humboldt Professorship Award, by the Deutsche Forschungsgemeinschaft (DFG, German Research Foundation) under Germany's Excellence Strategy - EXC-2111 - 390814868, Munich Center for Quantum Science and Technology, 
 Germany's Excellence Strategy  under Excellence Cluster Origins as well as ERC-AdG-2015 grant 694896.

\end{acknowledgments}

\section*{Appendix: Finding Parameter Scalings}\label{sec:app}

In order to determine how the rewriting values of $C_m$ and the corresponding rates $\Gamma$ scale with the parameter $X \in \lbrace N_c, \epsilon_m, C_0, \Delta N_c, K\rbrace$, the system has been time evolved with different $X$, with the remaining parameters fixed at the values given in \eqref{systemParameters}. For each $X$-value, the time evolutions have been 
done for many couplings $C_m \in [0,1]$ (or a larger interval), where we used a sampling step of $\delta C_m = 10^{-3}$ or smaller.

We have defined a rewriting value of $C_m$ as a value of $C_m$ for which the amplitude of $n_0$ exceeds the amplitude of $n_0$ in the free case, \ie the case of $C_m = 0$, by a factor of at least $1.2$.
For neighboring rewriting values (i.e., separated only by the increment $\delta C_m$), we only considered the one with the highest value of $\Gamma$. Around this value, we again performed time evolutions with a smaller sampling step of $\delta C_m = 5\cdot10^{-5}$. The reason why we did so 
is that the rate depends on $C_m$ very sensitively. Finally, we always selected the point with the highest rate.

In the following, we show the plots containing the data as well as fits for the individual scalings and further elaborate on our procedure.

\subsection*{$N_c$-Scaling}

\begin{figure*}
	\begin{subfigure}{0.45\textwidth}
		\includegraphics[width=\textwidth]{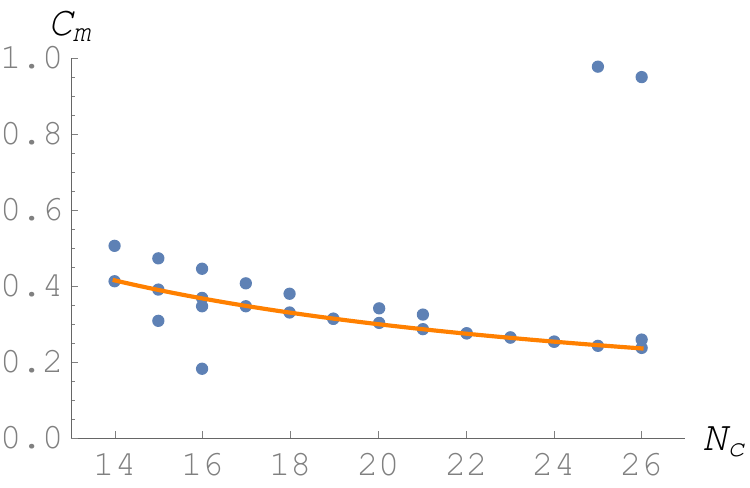}
		\caption{Rewriting values of $C_{\text{m}}$.} 
		\label{sfig:NDependenceCm}
	\end{subfigure}
	\hspace{0.05\textwidth}
	\begin{subfigure}{0.45\textwidth}
		\includegraphics[width=\textwidth]{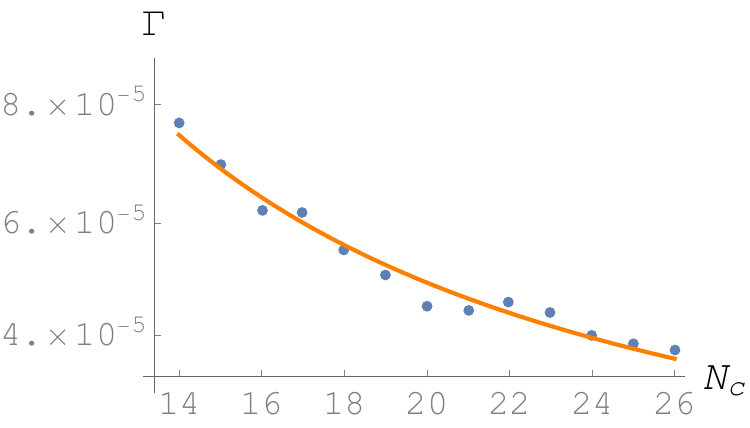}
		\caption{Rates $\Gamma$ at those rewriting values.}
		\label{sfig:NDependenceRate}
	\end{subfigure}
	\caption{Data and fits for the rewriting values of $C_m$ and the rates $\Gamma$ as function of $N_c$.
		$\Delta N_c$ has been varied to keep $N_c/\Delta N_c$ fixed. 
		}
	\label{fig:NDependence}
\end{figure*}

The data used to determine the scaling of $(C_m, \Gamma)$ with $N_c$ is shown in \fig \ref{fig:NDependence}.
The $N_c$-variation has been done varying also $\Delta N_c$, s.t. $N_c/\Delta N_c$ stays fixed.
The function fitted to the rewriting values is $f_{C}(N_c) = a \left( \frac{N_c}{22} \right)^{-b}$, with the fit result $a \approx 0.275$ and $b 
\approx 0.911$.
The function fitted to the rates is $f_{\Gamma}(N_c) = A \left( \frac{N_c}{22} \right)^{-B}$, with the fit result $A \approx 4.46 \cdot 10^{-5}$ 
and $B\approx 1.14$.

\subsection*{$\epsilon_m$-Scaling}

\begin{figure*}
	\begin{subfigure}{0.45\textwidth}
		\includegraphics[width=\textwidth]{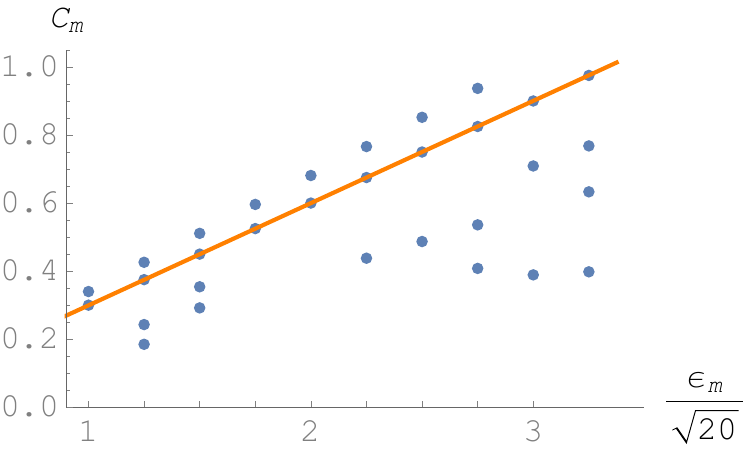}
		\caption{Rewriting values of $C_{\text{m}}$.} 
		\label{sfig:CgapDependenceCm}
	\end{subfigure}
	\hspace{0.05\textwidth}
	\begin{subfigure}{0.45\textwidth}
		\includegraphics[width=\textwidth]{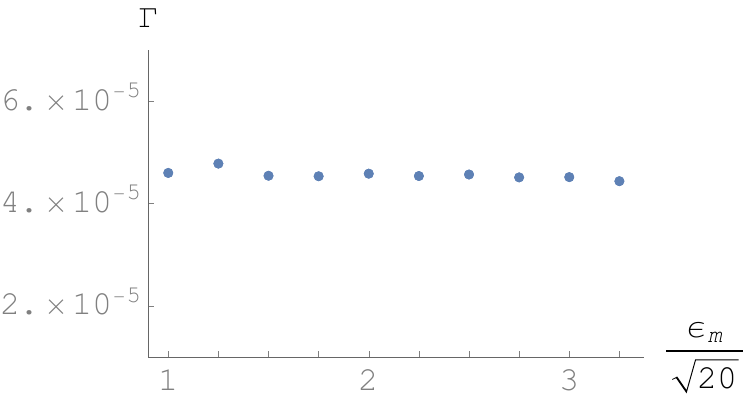}
		\caption{Rates $\Gamma$  at those rewriting values.}
		\label{sfig:CgapDependenceRate}
	\end{subfigure}
	\caption{Data and fit for the rewriting values of $C_m$ and the rates $\Gamma$ as function of $\epsilon_m$. 
	}
	\label{fig:CgapDependence}
\end{figure*}

The data used to determine the scaling of $(C_m, \Gamma)$ with $\epsilon_m$ is shown in \fig \ref{fig:CgapDependence}.
The function fitted to the rewriting values is $f_{C}(\epsilon_m) = a \epsilon_m$, with the fit result $a \approx 0.300$.
We observe that the scaling of the rate $\Gamma$ with $\epsilon_m$ is negligible compared to its scaling with other parameters.

\subsection*{$C_0$-Scaling}

\begin{figure*}
	\begin{subfigure}{0.45\textwidth}
		\includegraphics[width=\textwidth]{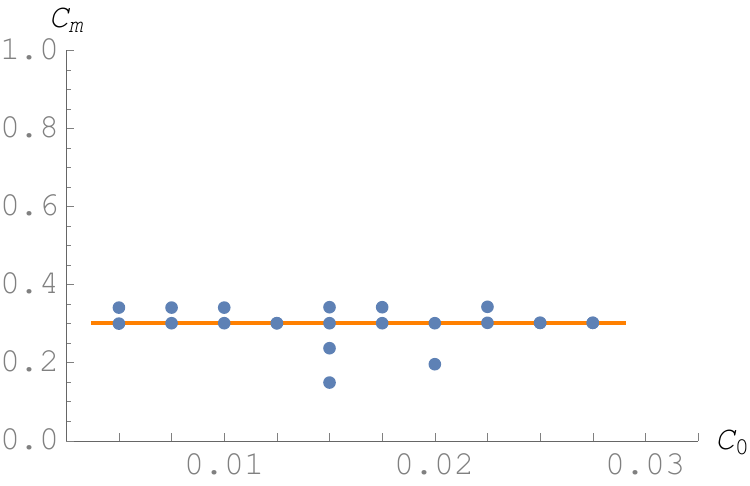}
		\caption{Rewriting values of $C_{\text{m}}$.} 
		\label{sfig:CconDependenceCm}
	\end{subfigure}
	\hspace{0.05\textwidth}
	\begin{subfigure}{0.45\textwidth}
		\includegraphics[width=\textwidth]{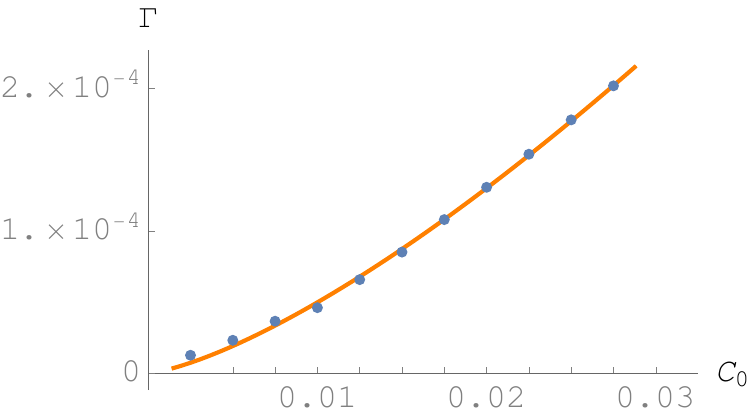}
		\caption{Rates $\Gamma$ at those rewriting values.}
		\label{sfig:CconDependenceRate}
	\end{subfigure}
	\caption{Data and fits for the rewriting values of $C_m$ and the rates $\Gamma$ as function of $C_0$. 
	}
	\label{fig:CconDependence}
\end{figure*}

The data used to determine the scaling of $(C_m, \Gamma)$ with $C_0$ is shown in \fig \ref{fig:CconDependence}.
For the scaling of the rewriting values with $C_0$, we observe that it is 
negligible compared to the scaling with other parameters.
The function fitted to the rates is $f_{\Gamma}(C_0) = A C_0^B$, with the fit result $A \approx 2.85 \cdot 10^{-2}$ and $B\approx 1.38$.

\subsection*{$\Delta N_c$-Scaling}

\begin{figure*}
	\begin{subfigure}{0.45\textwidth}
		\includegraphics[width=\textwidth]{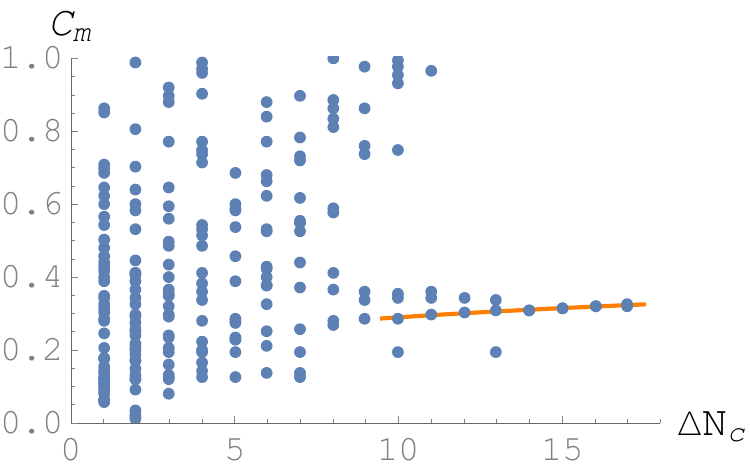}
		\caption{Rewriting values of $C_{\text{m}}$.} 
		\label{sfig:DeltaNDependenceCm}
	\end{subfigure}
		\hspace{0.05\textwidth}
		\begin{subfigure}{0.45\textwidth}
			\includegraphics[width=\textwidth]{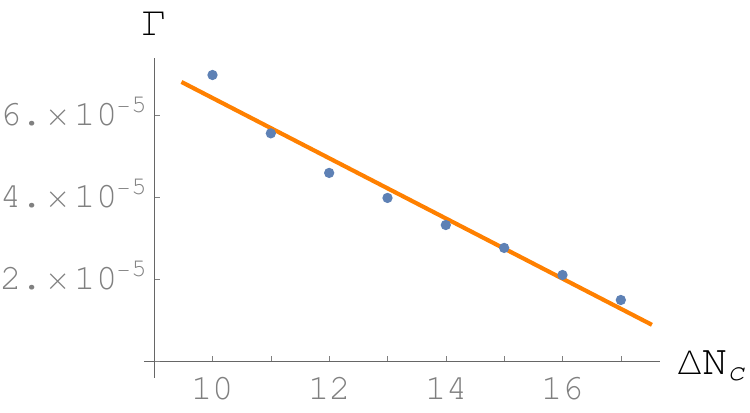}
			\caption{Rates $\Gamma$  at those rewriting values.}
			\label{sfig:DeltaNDependenceRate}
		\end{subfigure}
	\caption{Data and fits for the rewriting values of $C_m$ and the rates $\Gamma$ as function of $\Delta N_c$.}
	\label{fig:DeltaNDependence}
\end{figure*}

The data used to determine the scaling of $(C_m, \Gamma)$ with $\Delta N_c$ is shown in \fig \ref{fig:DeltaNDependence}.
The function fitted to the rewriting values is 
$f_{C}(\Delta N_c) = a \left( \frac{\Delta N_c}{12} \right)^{b}$, with the fit result $a \approx 0.300$ and $b\approx 0.207$.
\footnote
{A second scaling behavior seems to exist with $b\approx - 0.130$. Since $\Delta N_c/N_c \rightarrow 0$ in the limit of a large system, this scaling would be even less favorable for rewriting and we consequently do not consider it.}
The function fitted to the rates is $f_{\Gamma}(\Delta N_c) = A\left(1-B\frac{\Delta N_c}{20} \right)$, with the fit result $A \approx 1.38 \cdot 10^{-4}$ and $B\approx 1.07$.

\subsection*{$K$-Scaling}
When investigating how the system depends on $K$, the problem is that the 
size of the Hilbert space grows exponentially with the number of modes. Therefore, only values up to $K=8$ are numerically accessible. In order to be insensitive to effects of changing the relative occupation of the memory modes, we moreover restrict ourselves to $N_m=K/2$. Since the system appears to behave in a non-generic way for $K=2$, we are only left with three datasets corresponding to $K=4,6,8$. 

An additional difficulty arises from the fact that time evolutions for $K=8$ are already very time-consuming. Since many rewriting values exist at $K=8$, it is not feasible to perform fine scans, which are required for a precise determination of the rate, around each of them. To overcome 
this problem, we have selected a particular subset of rewriting values and only used them for the subsequent analysis. Namely, only those rewriting values of $C_m$ have been taken into account which have at least one neighboring rewriting value, i.e., a rewriting value separated solely by the increment $\delta C_m = 10^{-3}$. Of course, we have applied the same 
selection procedure for $K=4$ and $K=6$. Only the selected rewriting values are displayed in \fig \ref{fig:QDependence}, and only for those the rate has been determined by means of a finer scan.

\bibliography{cits3}
\include{extended2.bbl}

\end{document}